# Solvent/Antisolvent-Driven Etching of 2D Layered Halide Perovskite Microcrystals toward Endo-Epitaxial Heterostructures

*Ajeet Singh*[!*], *Sudhir Kumar Saini*[!!], *Hriday Kumar Pal*[!], *Cara Elisabeth Bradsher*[!,#], *Sirous Khabbaz Abkenar*[#], *Roman Krahne*[!!*], *Liberato Manna*[!*]

[!]*Nanochemistry, Istituto Italiano di Tecnologia, Genoa, Italy 16162*

[!!]*Optoelectronics, Istituto Italiano di Tecnologia, Genoa, Italy 16162*

[#]*Electron Microscopy Facility, Istituto Italiano di Tecnologia, Genoa, Italy 16162*

*Corresponding authors:*

*Ajeet Singh ( ajeet.singh@iit.it)*

*Roman Krahne ( roman.krahne@iit.it)*

*Liberato Manna ( liberato.manna@iit.it)*

## Abstract

Engineering internal interfaces within single crystals remains challenging, particularly in soft ionic materials such as layered halide perovskites (LHPs), for which conventional lithographic approaches are difficult to implement. Here we report a solvent/antisolvent-driven strategy for controlled etching of LHP microcrystals, producing well-defined rectangular cavities. Etching initiates at the top surface and proceeds preferentially along the crystallographic [100] and [010] directions. Cavity dimensions are tuned through etching time, temperature and etchant composition. The reaction can be interrupted and resumed, enabling stepwise cavity evolution through successive etching cycles. The resulting cavities serve as confined templates for regrowth of compositionally distinct LHPs, yielding endo-epitaxial heterostructures with designed internal architectures. The approach extends across diverse LHP compositions and antisolvent systems, while sequential precursor addition controls the extent of internal regrowth. This strategy establishes a route to engineering internal heterointerfaces in single-crystal perovskites through controlled chemical etching and confined growth.

**Introduction**

Two-dimensional (2D) LHPs have emerged as a versatile class of semiconductors owing to their natural quantum-well like electronic structure, high compositional tunability, and excellent optoelectronic properties.[1–4] These characteristics have enabled the development of diverse perovskite heterostructures, including vertical, lateral, and core/shell architectures, in which spatial control over composition and interfaces provides new opportunities for tailoring charge, exciton, and energy-transfer processes.[5–15] Beyond interfaces defined at the exterior of a crystal, however, the ability to deliberately engineer the surface architecture of a perovskite single crystal remains underexplored. This challenge is particularly relevant for soft, ionically bonded perovskites, for which conventional lithographic patterning and dry-etching approaches that work for robust semiconductor materials can be difficult to implement without compromising structural integrity or electronic behavior.[16–21] In other layered materials, the deliberate construction of internal crystal interfaces has emerged as a powerful route toward increasing structural complexity and accessing properties that are difficult to achieve in homogeneous crystals. In particular, endo-epitaxial heterostructures in layered transition-metal dichalcogenides demonstrate how selective modification and growth within a parent crystal can provide precise control over internal interfaces and composition.[17,22,23] Extending this concept to layered halide perovskites offers an attractive yet largely unexplored approach to creating three-dimensional architectures within an otherwise single-crystalline host. Such internal interfaces could provide new platforms for heterostructure formation while retaining the crystallographic continuity and structural definition of the parent crystal.

Recently, Dou and co-workers reported a solution-mediated approach for generating well-defined internal cavities in 2D perovskite microcrystals.[24] Exposure of the LHP microcrystals to isopropanol containing dissolved organic spacer cations induces selective chemical etching, producing square-shaped holes within the crystals. This work established that chemical modification of a perovskite surface can be translated into controlled 3D restructuring of the crystal interior, opening a pathway toward endo-structured perovskite architectures. However, the ability to systematically control the etching process and extend this strategy across different LHP compositions and processing conditions is so far highly restricted. Developing such control would provide new opportunities to engineer internal architectures in 2D perovskite single crystals and for exploiting them for subsequent heterostructure formation.

Here, we develop a solvent/antisolvent-driven chemical etching strategy for the controlled formation of surface-initiated cavities that extend into LHP microcrystals. Using acetonitrile as the solvent in combination with chlorobenzene, ethyl acetate, or diethyl ether as antisolvents, we achieve the reproducible formation of well-defined cavities in both Pb-based LHPs and Pb-free layered halide double perovskites while maintaining the single-crystalline character of the original crystals. Through comprehensive investigations of the obtained structures and the formation process, we elucidate the mechanism governing the solvent/antisolvent-driven etching process and the evolution of the holey crystal architecture. Complementary 3D kinetic Monte Carlo (kMC) simulations capture the time-dependent evolution of cavity morphology, providing insights into the coupled vertical and lateral expansion of the etched regions. We further exploit these patterned crystals as templates for selective endo-epitaxial growth, enabling the formation of $PEA_2PbI_4$–$PEA_2PbBr_4$ (where PEA = phenylethylammonium), $BA_2PbI_4$–$BA_2PbBr_4$ (where BA = butylammonium), and $PEA_2SnI_4$–$PEA_2PbBr_4$ heterostructures within the etched cavities. Our results establish chemical etching as a versatile route for crystal engineering in LHPs, demonstrating how surface-initiated cavities can serve as templates for constructing compositionally defined internal heterointerfaces and expanding the architectural design space of perovskite heterostructures.

**Results and Discussion**

$PEA_2PbBr_4$ microcrystals were chosen as the model system for the following experiments on account of their wide investigation and high control on the synthesis and reproducibility, unless stated otherwise.[5,6,25–27] **Figures 1a,b** provide the schematic illustration of the synthesis method and the solvent/antisolvent-driven etching strategy for preparing the etched microcrystals, respectively. The optical bright-field (BF) images of $PEA_2PbBr_4$ microcrystal show well-defined holes after etching for 1 min with a solution containing 150 μL acetonitrile and 350 μL chlorobenzene at room temperature. For detailed synthesis and etching reaction details, see the Experimental Section in the supplementary information (SI).

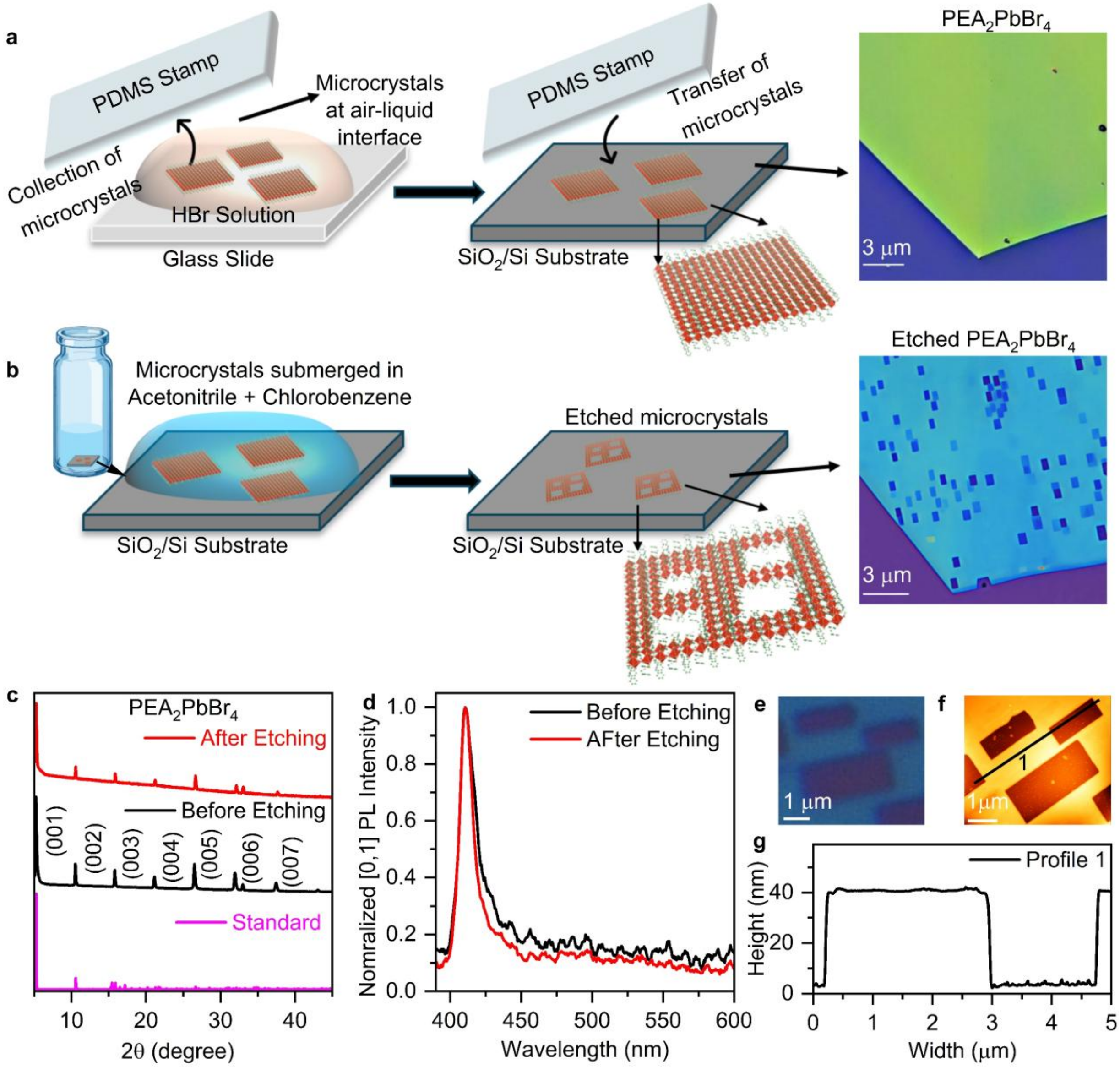

**Figure 1 | Solvent/antisolvent-driven etching of $PEA_2PbBr_4$ microcrystals. a** Schematic illustration of the synthesis and transfer of $PEA_2PbBr_4$ microcrystals and **b** their solvent/antisolvent-driven etching process. **c** XRD patterns of the microcrystals before and after etching, compared with the standard diffraction pattern (CCDC 754084). **d** PL spectra of the $PEA_2PbBr_4$ microcrystals before and after etching. **e** Cropped BF image of the etched

microcrystal surface and **f** the corresponding AFM image. **g** Height profile extracted from the AFM image along the line depicted in f.

**Figure 1c** provides the X-ray diffraction patterns (XRD) of as-synthesized $PEA_2PbBr_4$ microcrystals before and after etching, matched with the respective standard references.[28] The XRD patterns confirm that the crystal structure is not affected by the etching process and the etched microcrystals maintain the high intensity characteristic equidistant peaks along the (00l) direction. **Figure 1d** provides the photoluminescence (PL) spectra of $PEA_2PbBr_4$ before and after etching of the microcrystals, respectively. The PL spectra show no shift in the PL peak or any change in the peak-profile, confirming that the solvent/antisolvent-driven etching does not produce appreciable changes in the emission energy or spectral profile. **Figures 1e,f** provide an optical BF-image of a selected surface region of microcrystal and the corresponding atomic force microscopy (AFM) image. For the large area optical BF image and the corresponding AFM image of same microcrystal see **Figure S1**. The AFM image shows the sharp edges of the etched holes and their height profile, provided in **Figure 1g**, demonstrates that the surface of the unetched region of the microcrystal is flat, while for the bottom of the holes a certain roughness is evident.

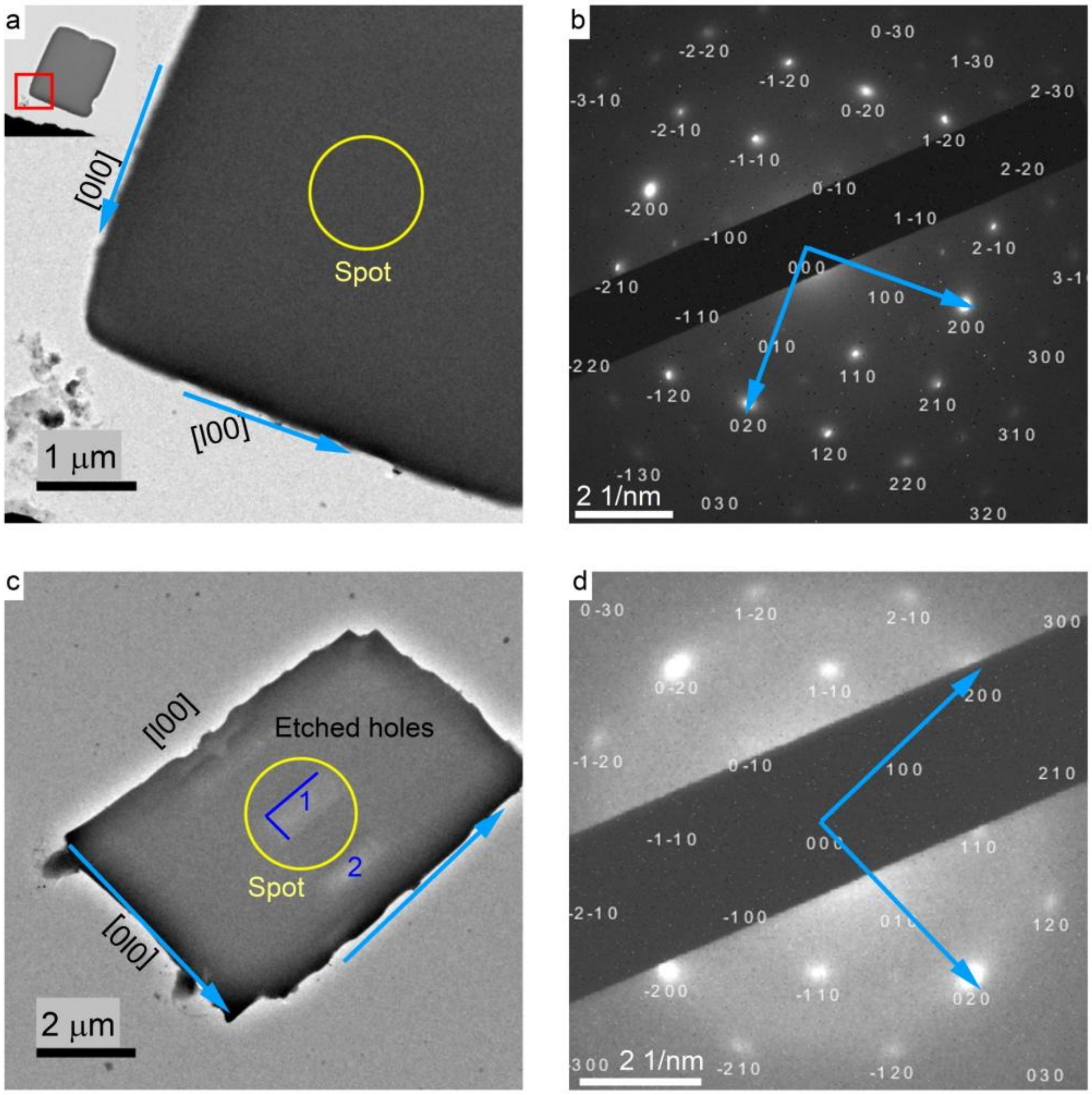

**Figure 2 | Crystallographic orientation of etched holes in $PEA_2PbBr_4$ microcrystals. a** Transmission electron microscopy (TEM) image of a $PEA_2PbBr_4$ microcrystal corner before etching, with the SAED collection spot circled in yellow. The inset shows a low-magnification TEM image of the entire microcrystal; the red square marks the corresponding corner. **b** Indexed SAED pattern acquired from the marked spot. **c** TEM image of the etched microcrystal showing two distinct holes (1 and 2). **d** Indexed SAED pattern acquired from the yellow-circled spot. The blue arrow indicates that the hole edges are parallel to the microcrystal edge.

To correlate the well-defined morphology of the etched cavities with the crystallographic orientation of the $PEA_2PbBr_4$ microcrystals, TEM combined with selected-area electron diffraction (SAED) was employed. **Figure 2a** presents a representative TEM image of a pristine $PEA_2PbBr_4$ microcrystal, with the inset showing the overall morphology of the microcrystal. The corresponding SAED pattern collected from the region marked by the circle in **Figure 2a** is shown in **Figure 2b**. The presence of well-defined, intense diffraction spots confirms the single-crystalline nature of the microcrystal. Indexing of the diffraction pattern further reveals that the crystal edges are oriented along the [100] and [010] crystallographic directions.[29] These directions are parallel to the corresponding crystal edges, as indicated by the blue arrows in **Figure 2a**, establishing the crystallographic orientation of the microcrystals prior to etching. The crystallographic relationship between the microcrystal morphology and the etched cavities was subsequently examined in the etched microcrystals. **Figure 2c** shows a TEM image of an etched $PEA_2PbBr_4$ microcrystal, in which two distinct cavities are identified as holes 1 and 2. The corresponding SAED pattern collected from the circled region of the etched microcrystal is presented in **Figure 2d**. The well-defined diffraction spots and their indexing confirm that the microcrystal retains its single-crystal nature after etching. Based on the indexed SAED pattern, the [100] and [010] directions are found to be parallel to the edges of the etched cavities, demonstrating that the cavity boundaries preferentially form along these crystallographic directions. Notably, the [100] and [010] directions are parallel to the external edges of the microcrystal, indicating that the internal cavity morphology follows the crystallographic symmetry of the parent microcrystal. The preferential formation of cavity edges along the [100] and [010] directions is consistent with the reported surface-energy landscape of $PEA_2PbBr_4$, where the {100} surfaces exhibit comparatively low surface energies and are therefore more stable.[5] Thus, the crystallographic selectivity observed for the etched cavity boundaries is consistent with the preferential stabilization of the low-energy {100} surfaces. In contrast, higher-energy surfaces are less thermodynamically favorable and may therefore be preferentially removed during the etching process, providing a possible origin for the observed crystallographic selectivity.

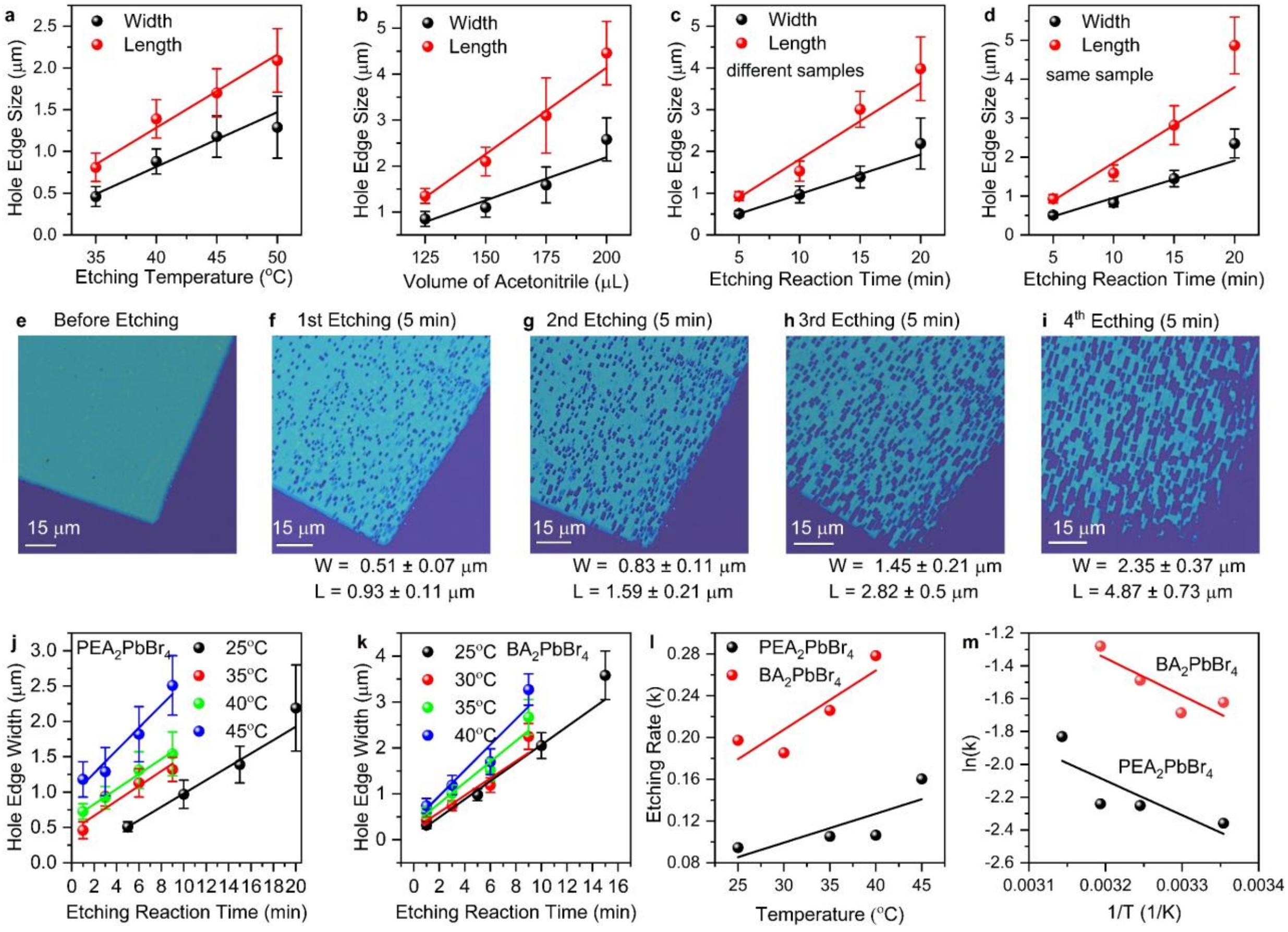

**Figure 3 | Control of hole formation through etching conditions and reaction cycling. a** Hole-etching dynamics at different temperatures using 100 μL acetonitrile and 400 μL chlorobenzene. **b** Hole-etching dynamics for different acetonitrile volumes, with the total etching-solution volume fixed at 500 μL. **c** Evolution of hole dimensions with etching time at room temperature. **d** Sequential hole-etching dynamics of the same microcrystal over successive 5 min etching cycles at room temperature. **e-i** BF images of the same microcrystal before etching and after successive 5 min etching cycles, respectively, corresponding to the etching dynamics shown in d. W and L denote the width and length of the etched holes, respectively. **j**,**k** Width of

the etched holes in $PEA_2PbBr_4$ and $BA_2PbBr_4$, respectively, as a function of etching time at different reaction temperatures. **l** Etching rates as a function of reaction temperature. **m** Arrhenius plot of $ln(k)$ versus $1/T$ used to determine the activation energy of the etching process. Error bars in **a-d**, **j**, and **k** represent the statistical variation in the hole dimensions.

To systematically investigate the etching dynamics, $PEA_2PbBr_4$ microcrystals were subjected to controlled etching under varying reaction temperatures, etching-solution compositions, and reaction durations. **Figure 3a** shows the average width and length of the etched holes formed after 1 min of etching using 100 µL acetonitrile and 400 µL chlorobenzene at different temperatures. Both dimensions increase linearly with temperature, indicating enhanced etching at elevated temperatures. The influence of etching-solution composition was subsequently investigated by varying the acetonitrile volume while maintaining a constant total volume of 500 µL. As shown in **Figure 3b**, the average hole dimensions increase linearly with increasing acetonitrile content. At a lower acetonitrile volume of 100 µL, cavity formation is relatively slow, requiring approximately 5 min to produce optically resolvable features. Conversely, increasing the acetonitrile volume beyond 250 µL accelerates dissolution to the extent that homogeneous removal of the crystal volume increasingly dominates over localized cavity formation, see **Figure S2**. These observations establish an intermediate acetonitrile-to-chlorobenzene ratio as essential for achieving controlled, localized etching while suppressing extensive dissolution. Representative BF optical images and corresponding hole-dimension distributions for the temperature- and etchant composition-dependent experiments are provided in **Figures S3** and **S4**, respectively.

The temporal evolution of the etched cavities was investigated at room temperature using an etching solution containing 100 µL acetonitrile and 400 µL chlorobenzene. As shown in **Figure 3c**, both the width and length of the etched holes increase linearly with reaction time over the investigated 20 min period, indicating an approximately constant rate of cavity expansion. To investigate whether etching could be interrupted and subsequently resumed, the same microcrystal was subjected to successive 5 min etching cycles, with intermediate removal from the solution for optical imaging before reimmersion in fresh etching solution. As shown in **Figure 3d–i**, the cumulative evolution of the cavity dimensions closely follows that observed during continuous etching over the corresponding total reaction time. This demonstrates that the etching process can be temporarily interrupted and subsequently resumed at pre-existing cavities without appreciably altering their subsequent growth trajectory. Continued etching at existing sites occurs alongside the initiation of new cavities, demonstrating the reproducibility of the process over successive cycles. The corresponding BF images and hole-dimension distributions for the time-dependent and cyclic etching experiments are provided in **Figures S5** and **S6**, respectively.

To quantify the temperature dependence of the etching kinetics and compare the influence of organic spacer chemistry, the apparent activation energies were determined for $PEA_2PbBr_4$ and $BA_2PbBr_4$ using the same acetonitrile/chlorobenzene composition (100/400 μL). The corresponding BF optical images and hole-dimension distributions for $PEA_2PbBr_4$ and $BA_2PbBr_4$ at the different reaction temperatures are provided in **Figures S7–S9** and **Figures S10–S13**, respectively. **Figure 3j,k** show the temporal evolution of the average hole width at different temperatures for $PEA_2PbBr_4$ and $BA_2PbBr_4$, respectively. The approximately linear increase in hole width with time at each temperature allows the corresponding etching rates to be extracted from the slopes. As shown in **Figure 3l**, the etching rates increase systematically with temperature for both compositions. The temperature dependence was analyzed using the Arrhenius relationship, $ln(k) = ln(A) - E_a/RT$, where $k$ is the etching rate, $A$ is the pre-exponential factor, $E_a$ is the apparent activation energy, $R$ is the gas constant, and $T$ is the absolute temperature.[24] The corresponding Arrhenius plots are presented in **Figure 3m**, from which apparent activation energies of 17.54 kJ mol$^{-1}$ for $PEA_2PbBr_4$ and 18.94 kJ mol$^{-1}$ for $BA_2PbBr_4$ were obtained. Despite these comparable activation energies, the two compositions exhibit distinct absolute etching rates. Well-defined cavities were readily observed in $BA_2PbBr_4$ after 1 min of etching at room temperature, whereas optically resolvable cavities were not observed in $PEA_2PbBr_4$ under the same conditions. This difference indicates that the organic spacer chemistry influences the overall kinetics of cavity formation, potentially through differences in spacer-cation removal and interactions with the etching medium. Thus, although both compositions exhibit similar temperature-dependent kinetic behavior, their absolute etching rates are influenced by the chemical nature of the organic spacer.

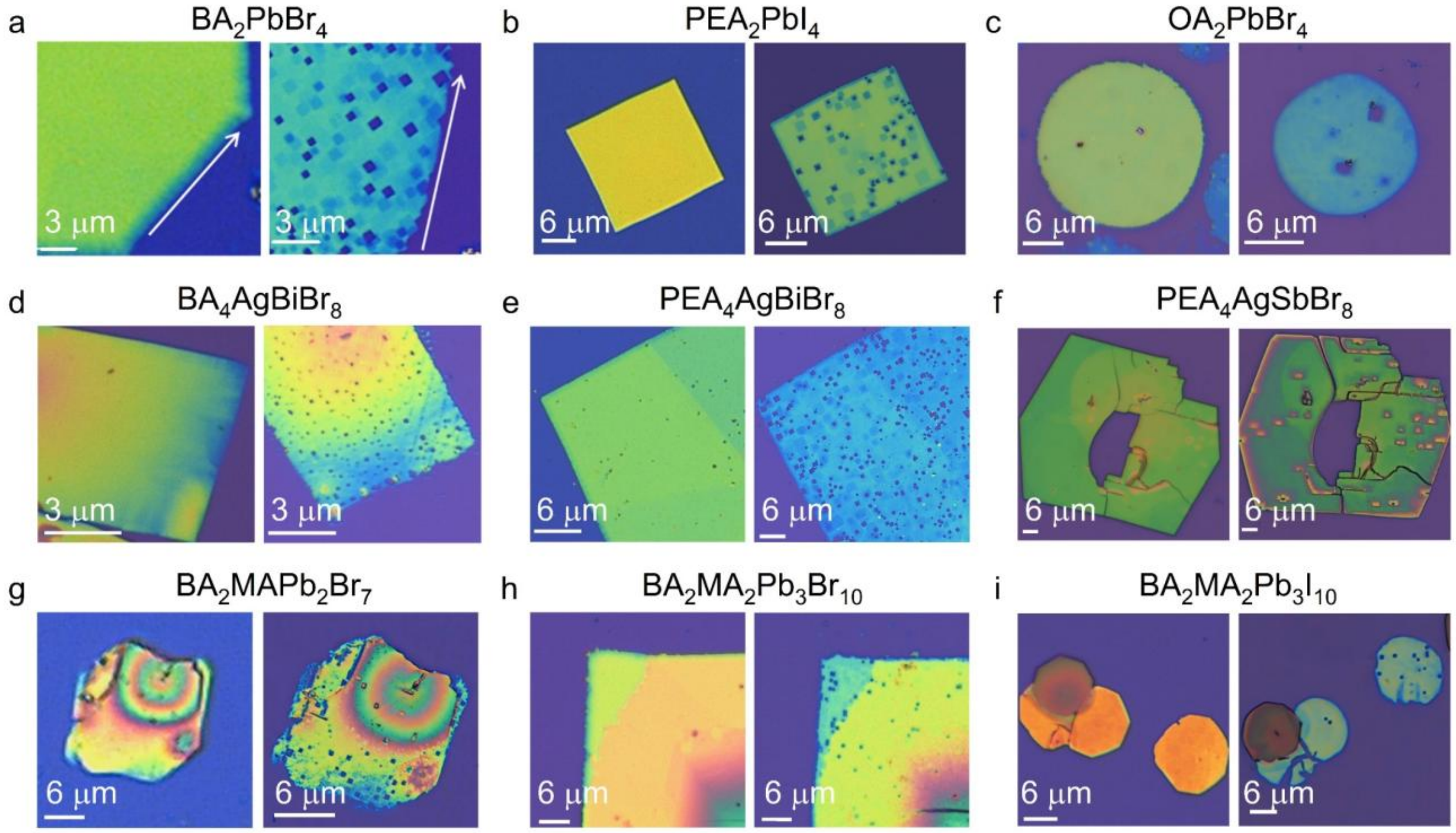

**Figure 4 | Compositional scope of solvent/antisolvent-driven etching across layered halide perovskites.** Optical BF images of pre-etched and after etching microcrystals. **a** $BA_2PbBr_4$, **b**

$PEA_2PbI_4$, **c** $OA_2PbBr_4$, **d** $BA_4AgBiBr_8$, **e** $PEA_4AgBiBr_8$, **f** $PEA_4AgSbBr_8$, **g** $BA_2MAPb_2Br_7$, **h** $BA_2MA_2Pb_3Br_{10}$, and **i** $BA_2MA_2Pb_3I_{10}$, respectively.

To assess the compositional scope of our solvent/antisolvent-driven etching strategy, we extended the approach to a diverse range of LHPs. **Figure 4** presents optical BF images of pristine and etched microcrystals encompassing Pb-based LHPs, including $BA_2PbBr_4$, $PEA_2PbI_4$, $OA_2PbBr_4$ (where OA = octylammonium); Pb-free layered halide double perovskites, including $BA_4AgBiBr_8$, $PEA_4AgBiBr_8$, $PEA_4AgSbBr_8$; and higher-n methylammonium (MA)-based LHPs, including $BA_2MAPb_2Br_7$ (n = 2), $BA_2MA_2Pb_3Br_{10}$ (n = 3), and $BA_2MA_2Pb_3I_{10}$ (n = 3). In each case, well-defined internal cavities could be generated by appropriately tuning the acetonitrile/chlorobenzene composition and etching duration at room temperature. The hole size distribution plots and the XRD patterns are provided in **Figure S14**, and **Figure S15**, respectively, while the etching conditions used for each composition are summarized in **Table S2**. Although internal cavity formation was achieved across all investigated compositions, the optimal etching conditions varied substantially with the inorganic framework and organic spacer chemistry. In particular, the Pb-free double perovskites exhibited considerably faster dissolution kinetics than their Pb-based counterparts. For example, $BA_4AgBiBr_8$ microcrystals underwent complete dissolution within 10 seconds when exposed to an etching solution containing only 10 μL acetonitrile and 490 μL chlorobenzene. In contrast, complete dissolution of $BA_2PbBr_4$ under the same duration and temperature required substantially higher acetonitrile fractions (>250 μL). Similarly, $PEA_4AgBiBr_8$ and $PEA_4AgSbBr_8$ microcrystals were completely dissolved within 120 s using 200 μL acetonitrile and 300 μL chlorobenzene, whereas their Pb-based PEA analogues required substantially higher acetonitrile fractions (>300 μL for the bromide composition) to achieve comparable dissolution under the investigated conditions. These observations indicate that the inorganic framework contributes to the susceptibility of layered perovskites to solvent-mediated dissolution. Unlike the corner-sharing $[PbX_6]^{4-}$ octahedral framework in conventional Pb-based LHPs, the double perovskites contain alternating corner-sharing $[AgX_6]^{5-}$ and $[BiX_6]^{3-}$ (or $[SbX_6]^{3-}$ octahedra. The chemically distinct metal–halide coordination environments, differences in Ag–Br and Bi–Br bond lengths, and reported tetragonal distortions at the $[AgBr_6]^{5-}$ sites introduce structural heterogeneity within the inorganic layers, which may influence the energetics of ion removal.[30,31] These features could therefore contribute to the enhanced dissolution susceptibility observed for the double perovskites. However, dissolution also depends on the interactions of the released metal and halide ions with the etching solution, including its ability to solvate or complex different metal cations.[32] Consequently, the relative contributions of metal–halide bonding, octahedral distortions, and solvent–ion interactions to the enhanced dissolution remain to be established.[29],[30]

We further examined whether the choice of antisolvent is restricted to chlorobenzene. Ethyl acetate, diethyl ether, and n-butanol were therefore evaluated as alternative antisolvents using $PEA_2PbBr_4$ microcrystals. All three antisolvents enabled the formation of well-defined cavities under appropriately adjusted etching conditions. **Figure S16** shows the optical BF images and corresponding hole-size distributions of the pristine and etched microcrystals obtained with the different antisolvents at room temperature. The etchant compositions and corresponding etching durations are summarized in **Table S3**. Notably, the formation of cavities was observed across all four antisolvents examined, indicating that the etching process is not restricted to chlorobenzene as the antisolvent.

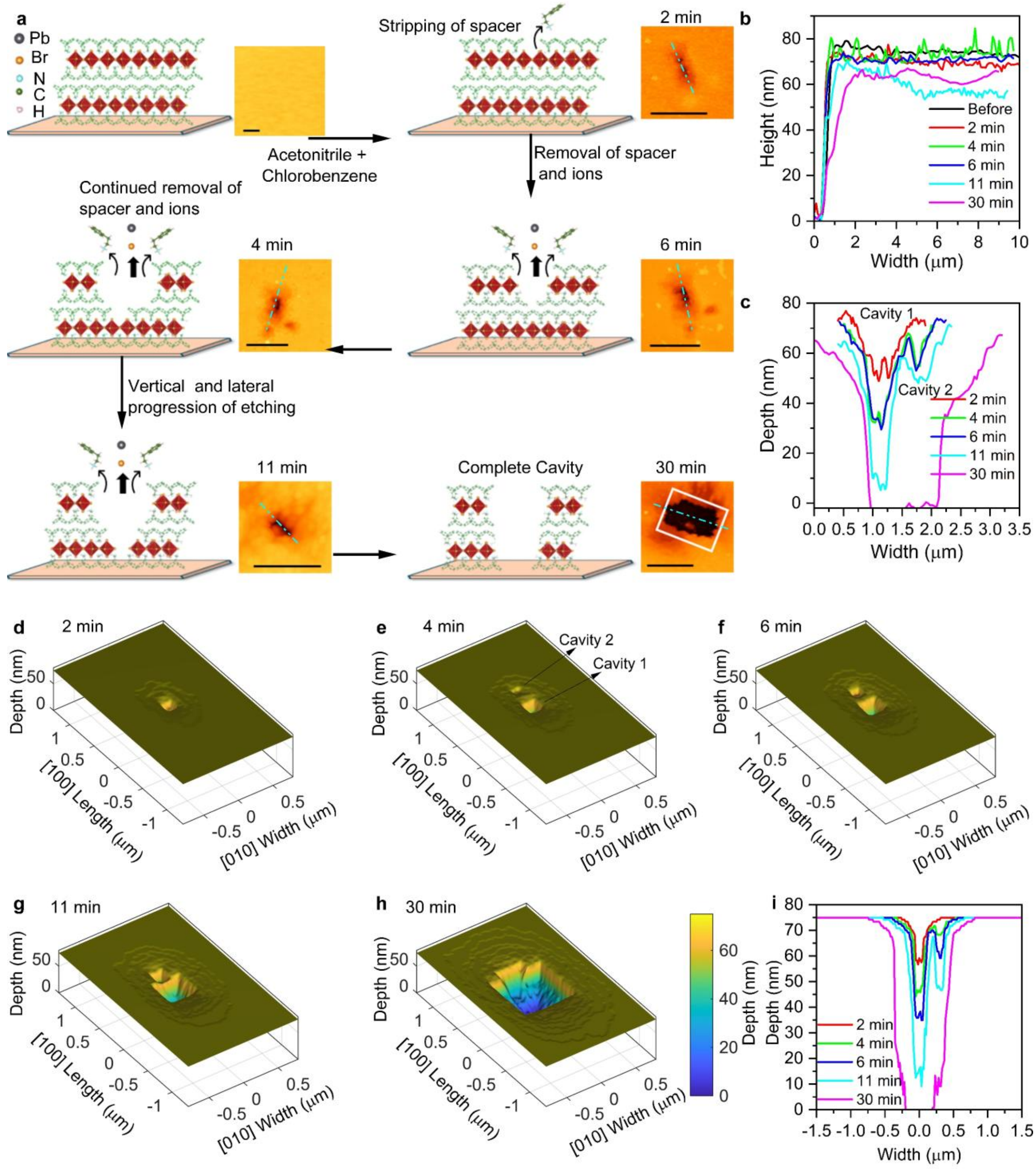

**Figure 5 | Evolution of etched cavities and their three-dimensional morphological reconstruction. a** Schematic graphical illustration of etching mechanism in $PEA_2PbBr_4$, the cropped-out AFM image shows the evolution of a selected cavity with successive etching cycles.

**b** The height profile of the $PEA_2PbBr_4$ microcrystal after etching for different times in successive etching cycles. **c** The depth profile of the etched cavity from the AFM image in b along the white line, after successive etching cycles. Scale bar in AFM images is 1 μm and the depth profiles were extracted along the white line in the AFM images. **d-h** Reconstructed 3D cavity morphologies obtained from the kMC simulations at etching times of 2, 4, 6, 11 and 30 min, respectively. The color scale represents the local cavity depth. The two simulated etching regions, Cavity 1 and Cavity 2, are indicated in **e**. **i** Simulated cross-sectional depth profiles showing the evolution of cavity depth and lateral dimensions with increasing etching time. The profiles illustrate the progressive deepening and lateral expansion of the cavities.

The etching process consistently initiates from the top surface of the microcrystals rather than from the side facets or edges, irrespective of microcrystal thickness, lateral dimensions, or the identity of the metal or organic spacer cation, as evidenced by optical microscopy. This points to a surface-controlled process rather than an edge-mediated dissolution pathway, and excludes edge-localized defects as the dominant initiation sites. We therefore propose a self-propagating, layer-by-layer topochemical dissolution mechanism that initiates at a surface instability or defect. [14] Then the etching front propagates vertically through the alternating organic and inorganic layers, generating crystallographically defined cavities that progressively deepen into the crystal while simultaneously expanding laterally during repeated etching cycles The exposed inorganic framework subsequently undergoes dissolution through the removal of metal and halide ions, enabling the repeated dissolution of organic and inorganic layers and thereby allowing the etching front to progressively propagate into the crystal (see illustration in **Figure 5a**). To distinguish such directional propagation from homogeneous dissolution, we monitored the evolution of etched cavities in $PEA_2PbBr_4$ microcrystals in successive etching cycles using a 100 µL acetonitrile and 400 µL chlorobenzene solution at room temperature. Cropped AFM images of a representative etched cavity at different stages of successive etching are shown in **Figure 5a**, while large-area, high-resolution AFM images and the corresponding optical BF images acquired after each successive etching cycle are provided in **Figure S17-S22**. **Figure 5b** shows that the height of the microcrystal following successive etching cycles remains largely preserved during the initial stages, evidencing the localized nature of the etching process. The extracted depth profiles provided in **Figure 5c**, for the cavity monitored in **Figure 5a** show initially a shallow, conical morphology that progressively increases in both depth and lateral dimensions with successive etching cycles. The continued vertical propagation into the crystal at the pre-existing cavity underpins the importance of the reaction sites on the surface. The micrometer-scale lateral expansion occurs preferentially along the crystal plane. With increasing etching time, the cavity therefore evolves from an initially shallow, conical depression into a deeper cavity with increasingly steep walls, while the surrounding microcrystal remains largely preserved.

To investigate the spatiotemporal evolution of the etched cavities, we developed a three-dimensional n-fold-way (Bortz–Kalos–Lebowitz) kMC model algorithm.[33,34] The model represents the layered perovskite film as a three-dimensional lattice, with the in-plane directions corresponding to the crystallographic [100] and [010] directions and the vertical direction representing the crystal depth with respect to the surface. Material removal is governed by two effective processes: lateral dissolution at exposed step edges and vertical dissolution within a localized etching channel. The lateral dissolution rate is anisotropic for the $PEA_2PbBr_4$ microcrystals, with the rate along the [100] direction 1.8 times larger (based on our experimental data from the aspect-average ratio of cavity length to width) than that along the [010] one.

Furthermore, a depth-dependent reduction in the lateral dissolution rate generates sloped cavity walls, while a time-dependent vertical etching channel allows progressive penetration into the underlying crystal.

The simulated three-dimensional morphologies reveal the progressive development of localized etching sites into extended cavities, see **Figure 5d-h**. At 2 min, the simulated cavity morphology features a relatively shallow and spatially confined depression. At 4 min, two distinct etching regions (Cavities 1 and 2) are discernible in the reconstructed morphology. Continued etching results in the progressive development of these features, with increasing penetration depth and lateral propagation along the crystal surface. At 6 and 11 min, the cavities exhibit increasingly pronounced three-dimensional profiles, reflecting the combined contributions of lateral edge dissolution and vertical material removal. By 30 min, the simulation produces a substantially developed cavity morphology, with extensive lateral dimensions and penetration reaching the bottom of the 75 nm film.

This evolution of the cavity dimensions corresponds well to our experimental observation as detailed in **Table S4** in the **SI**. The simulated cross-sectional profiles reported in **Figure 5i**, provide additional insight into the development of cavity geometry. To model the progressive narrowing of the cavity with increasing depth we imposed a depth dependence of lateral dissolution in the model, which produces more shallow-sloped upper walls and steeper lower walls. This geometrical evolution is consistent with the experimentally observed cavity profiles, although the simulated upper walls and deeper shaft remain narrower than those measured experimentally.

To reproduce the experimentally observed appearance of two adjacent cavities and their eventual coalescence, a second etching site was incorporated into the simulation along [100] crystal direction. The two cavities were treated as independently evolving regions, with their combined morphology represented by the union of their respective etched volumes. Their interaction was therefore approximated geometrically through the merging of their spatial extents rather than through a modification of their individual dissolution kinetics. The resulting morphology captures the development of an extended merged cavity at longer etching times. Overall, the 3D kMC simulations provide a spatially resolved representation of the evolution of the etched cavities, capturing the principal trends in vertical penetration, anisotropic lateral expansion and the development of complex cavity geometries. The simulations therefore provide a phenomenological framework for describing cavity evolution and identifying the contributions of anisotropic lateral propagation and vertical etching. Interestingly, the depth-dependent decrease in lateral dissolution incorporated into the kMC model may reflect the accumulation of dissolved species within the confined cavity that reduces the vertical etching rate, favoring the formation of wider openings and narrower, steeper cavity walls.

To demonstrate spatially confined etching, we coated $PEA_2PbBr_4$ microcrystals with a patterned polymethyl methacrylate (PMMA) film. **Figure S23a,b** show the optical BF images of square hole patterned PMMA coated on top of $PEA_2PbBr_4$ microcrystal before and after etching for 30 sec with 200 μL acetonitrile and 300 μL chlorobenzene solution, respectively. **Figure S23c** shows the BF image of the microcrystal after washing with chlorobenzene, and complete removal of the PMMA coating. **Figure S23d** shows the AFM image of the selected area and **Figure S23e** provides the extracted height profiles along the line numbered 1 in AFM image. The AFM analysis reveals that pristine microcrystal has a thickness of approximately 640 nm, whereas etching for only 30 seconds produces holes on the surface with depths exceeding 200 nm. From both profiles it is established that the etching propagates preferentially in vertical direction and the lateral expansion of the cavity is confined by the PMMA coating, demonstrating successful spatial confinement.

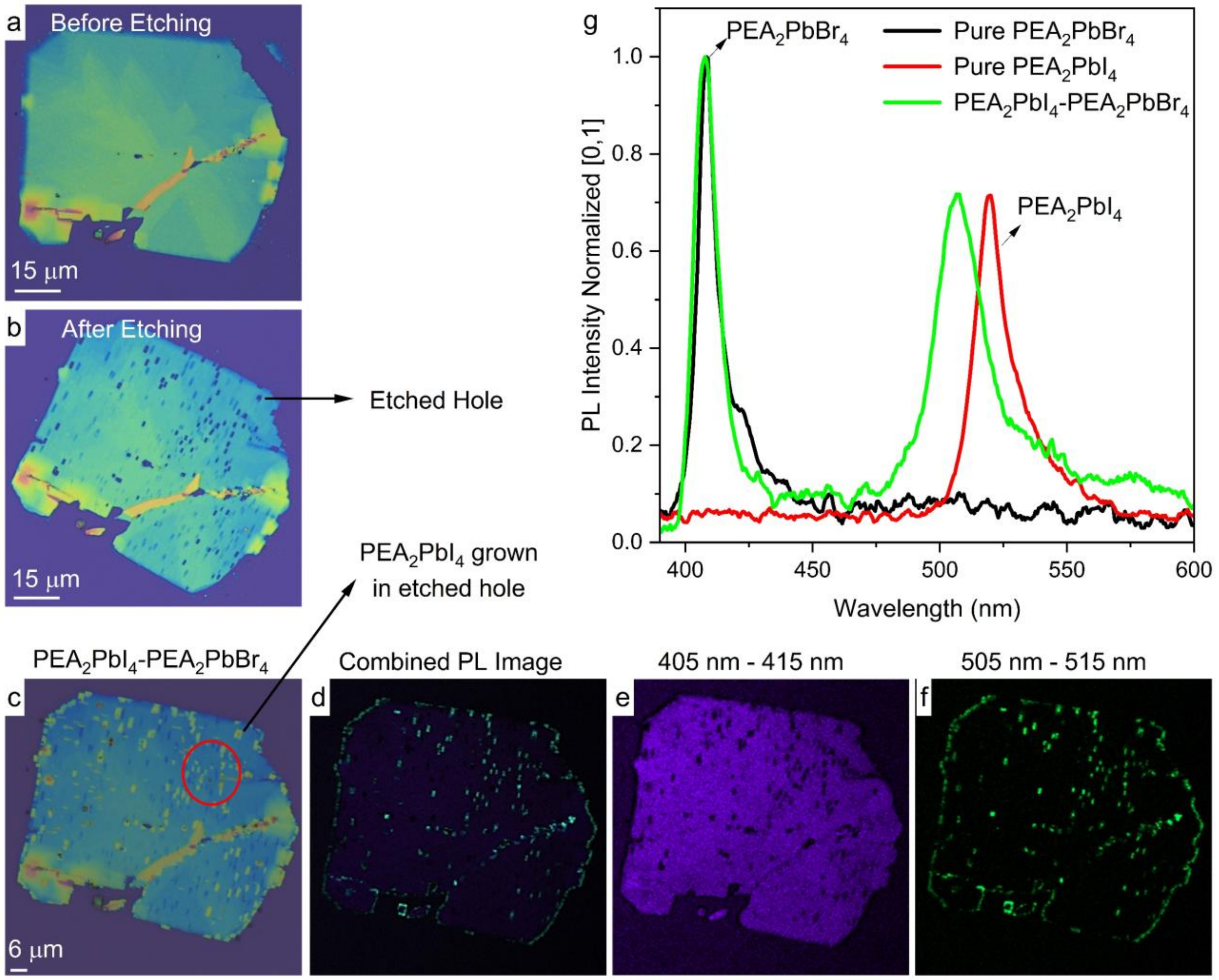

**Figure 6 | Cavity-confined growth of compositionally distinct perovskite heterostructures.** **a** BF image of pristine $PEA_2PbBr_4$ microcrystal. **b** BF image of the same microcrystal after etching. **c** BF image of $PEA_2PbI_4$-$PEA_2PbBr_4$ heterostructure microcrystal. **d** Combined all channel confocal PL image of $PEA_2PbI_4$-$PEA_2PbBr_4$. **e** Extracted 405 nm - 415 nm channel. **f** Extracted 505 nm - 515 nm channel. **g** The PL spectra of etched $PEA_2PbBr_4$ microcrystal and the $PEA_2PbI_4$-$PEA_2PbBr_4$ HS collected from the circled spot in HS BF image in **c**.

Following the successful demonstration of controlled internal interface formation through chemical etching of microcrystals, we synthesized $PEA_2PbI_4$–$PEA_2PbBr_4$ endo-epitaxial heterostructures (for synthesis details see Experimental Section in the SI). **Figure 6a,b** show the optical BF images of a pristine and chemically etched $PEA_2PbBr_4$ microcrystal, respectively. The etching was performed using 150 μL acetonitrile and 350 μL chlorobenzene for 2 min at room temperature. **Figure 6c,d** present the BF and corresponding PL images, respectively, of the resulting $PEA_2PbI_4$–$PEA_2PbBr_4$ heterostructure, with the PL image obtained by combining all confocal emission channels. **Figure 6e** shows the 405–415 nm emission channel, revealing purple emission from the remaining $PEA_2PbBr_4$ regions, while no detectable emission is observed from the etched regions. In contrast, the 505–515 nm channel shown in **Figure 6f** exhibits green emission originating from the $PEA_2PbI_4$-filled regions. **Figure 6g** compares the PL spectra recorded from the etched $PEA_2PbBr_4$ microcrystal and the $PEA_2PbI_4$–$PEA_2PbBr_4$ heterostructure, with the later collected from the circular region marked in **Figure 6c**. Following heterostructure formation, the PL peak of the newly grown $PEA_2PbI_4$ region was observed to be blue-shifted by approximately 13 nm relative to that of pristine $PEA_2PbI_4$ microcrystals. This shift is attributed to Br alloying in the iodide phase occurring during the heterostructure synthesis.[4] The optical microscopy results confirm the successful formation of endo-epitaxial $PEA_2PbI_{4-x}$–$PEA_2PbBr_4$ heterostructures within the etched microcrystals. **Figure S24** presents the XRD patterns of the heterostructure alongside those of the pristine and etched microcrystals. No additional diffraction peaks or measurable peak shifts were observed upon heterostructure formation, consistent with previously reported observations for related lateral heterostructures.[5,24] To further investigate the formation and growth of the iodide phase within the etched regions, we synthesized heterostructures with a larger $PEA_2PbI_4$ growth volume. The corresponding optical BF and PL images are provided in **Figure S25**. The heterostructure with a larger $PEA_2PbI_4$ region exhibits a substantially stronger green PL emission, with the PL peak shifted by only approximately 5 nm relative to pristine $PEA_2PbI_4$ microcrystals. The relatively small shift and the absence of a distinct additional emission peak suggest that the newly grown region consists predominantly of $PEA_2PbI_4$, with only a small contribution from Br. The larger $PEA_2PbI_4$ regions were obtained by repeatedly adding the precursor solution onto the same etched microcrystals in multiple cycles at 35 °C. This demonstrates that the dimensions of the newly grown perovskite can be systematically controlled by varying the number of precursor-addition cycles. The elemental distribution maps obtained by scanning electron microscopy–energy-dispersive X-ray spectroscopy (SEM–EDS), together with the corresponding optical BF image of the $PEA_2PbI_4$–$PEA_2PbBr_4$ heterostructure, are provided in **Figure S26**, while the corresponding summed EDS spectrum is provided in **Figure S27**. The Pb and Br elemental maps show that both elements are distributed throughout the microcrystal, whereas I is predominantly localized within the newly grown $PEA_2PbI_4$ region, providing further evidence for the spatially selective formation of $PEA_2PbI_4$ within the etched regions. To examine the applicability of the etching-assisted growth strategy to different LHP compositions, we extended the approach to the synthesis of $BA_2PbI_4$–$BA_2PbBr_4$ and $PEA_2SnI_4$–$PEA_2PbBr_4$ endo-epitaxial heterostructures. The corresponding optical BF images, together with the PL spectra, are provided in **Figure S28a-d**, and **S28e-h**, respectively. In both cases, the newly grown perovskite occupies the internally etched regions, while the optical contrast enables the newly formed heterostructure regions to be distinguished from the parent microcrystals. These results demonstrate that the approach can be extended beyond the $PEA_2PbI_4$–$PEA_2PbBr_4$ system to other combinations of organic spacer cations and metal/halide compositions.

## Conclusion

We have established solvent/antisolvent-driven chemical etching as a versatile strategy for engineering internal architectures in LHP microcrystals. The process enables the controlled formation of well-defined surface cavities, whose dimensions and morphological evolution can be systematically modulated through etchant composition, reaction time, and temperature. Importantly, terminating the etching reaction arrests further morphological evolution, providing temporal control over cavity formation. The process can also be resumed after termination, enabling stepwise expansion and evolution of the same internal cavities over multiple etching cycles. Its reproducibility across different LHP compositions and antisolvent systems highlights the broad applicability of this approach. Beyond cavity formation, we demonstrate that these chemically defined voids can serve as confined templates for subsequent epitaxial growth, enabling endo-epitaxial heterostructures, including $PEA_2PbI_4$–$PEA_2PbBr_4$, $BA_2PbI_4$–$BA_2PbBr_4$, and $PEA_2SnI_4$–$PEA_2PbBr_4$. Furthermore, sequential precursor addition provides an additional means of controlling the dimensions of the epitaxially regrown phase. Collectively, this etching–regrowth strategy establishes a route to constructing spatially defined, internal heterointerfaces within single-crystal perovskites, extending crystal engineering beyond conventional surface-defined architectures. Looking ahead, further control over cavity geometry and crystallographic orientation, together with the integration of multiple compositions through sequential etching and regrowth, could enable increasingly complex three-dimensional heterostructures. Such architectures also offer opportunities to investigate how confined internal interfaces influence excitonic, charge-transport, and optoelectronic properties, opening new directions for the design of functional perovskite materials.

## Acknowledgement

A.S. acknowledges funding from Horizon Europe through the MSCA-PF agreement no. 101150329 (TWISTPVSK). H. K. P., C. E. B., and L. M. acknowledge funding from European Research Council through the ERC Advanced Grant NEHA (grant agreement n. 101095974). S.K.A. acknowledge Electron Microscopy Facility. R.K. acknowledge funding Project 101131111 – DELIGHT.

## Ethics Declaration

Authors declare no conflict of interest.

## References

1. Stoumpos, C. C. *et al.* Ruddlesden–Popper Hybrid Lead Iodide Perovskite 2D Homologous Semiconductors. *Chem. Mater.* **28**, 2852–2867 (2016).
2. Mao, L., Stoumpos, C. C. & Kanatzidis, M. G. Two-Dimensional Hybrid Halide Perovskites: Principles and Promises. *J. Am. Chem. Soc.* **141**, 1171–1190 (2019).
3. Li, X., Hoffman, J. M. & Kanatzidis, M. G. The 2D Halide Perovskite Rulebook: How the Spacer Influences Everything from the Structure to Optoelectronic Device Efficiency. *Chem. Rev.* **121**, 2230–2291 (2021).
4. Blancon, J.-C., Even, J., Stoumpos, C. C., Kanatzidis, M. G. & Mohite, A. D. Semiconductor physics of organic–inorganic 2D halide perovskites. *Nat. Nanotechnol.* **15**, 969–985 (2020).
5. Shi, E. *et al.* Two-dimensional halide perovskite lateral epitaxial heterostructures. *Nature* **580**, 614–620 (2020).

6. Singh, A. *et al.* Two-Dimensional Halide Pb-Perovskite–Double Perovskite Epitaxial Heterostructures. *J. Am. Chem. Soc.* **145**, 19885–19893 (2023).
7. Xia, M. *et al.* Kinetic Wulff-shaped heteroepitaxy of phase-pure 2D perovskite heterostructures with deterministic slab thickness. *Nat. Synth.* **4**, 380–390 (2025).
8. Tatarinov, D. A., Schleusener, A. & Krahne, R. Controlling Energy Flow in Perovskite Heterostructures Through Dimensionality and Phase Engineering. *Adv. Sci.* **12**, e05971 (2025).
9. Akriti *et al.* Layer-by-layer anionic diffusion in two-dimensional halide perovskite vertical heterostructures. *Nat. Nanotechnol.* **16**, 584–591 (2021).
10. Fu, Y. *et al.* Multicolor Heterostructures of Two-Dimensional Layered Halide Perovskites that Show Interlayer Energy Transfer. *J. Am. Chem. Soc.* **140**, 15675–15683 (2018).
11. Lu, Y. *et al.* Layer-by-layer epitaxial growth of perovskite heterostructures with tunable band offsets. *Science* **390**, 716–721 (2025).
12. Pan, D. *et al.* Deterministic fabrication of arbitrary vertical heterostructures of two-dimensional Ruddlesden–Popper halide perovskites. *Nat. Nanotechnol.* **16**, 159–165 (2021).
13. Filippi, U. *et al.* Deterministic Nucleation of Nanocrystal Superlattices on 2D Perovskites for Light-Funneling Heterostructures. *Adv. Mater.* **n/a**, e74290.
14. Schleusener, A. *et al.* Heterostructures via a Solution-Based Anion Exchange in Microcrystalline 2D Layered Metal-Halide Perovskites. *Adv. Mater.* **36**, 2402924 (2024).
15. Faraji, M. *et al.* Heterostructure Design in Two-Dimensional Perovskites by Sequential Recrystallization. *Adv. Mater.* **n/a**, e75593.
16. Zhang, Y., Zhu, S., Hu, J. & Gu, M. Femtosecond laser direct nanolithography of perovskite hydration for temporally programmable holograms. *Nat. Commun.* **15**, 6661 (2024).

17. Zhang, Z. *et al.* Endoepitaxial growth of monolayer mosaic heterostructures. *Nat. Nanotechnol.* **17**, 493–499 (2022).
18. Lin, C.-H. *et al.* Orthogonal Lithography for Halide Perovskite Optoelectronic Nanodevices. *ACS Nano* **13**, 1168–1176 (2018).
19. Zhang, P., Yang, G., Li, F., Shi, J. & Zhong, H. Direct in situ photolithography of perovskite quantum dots based on photocatalysis of lead bromide complexes. *Nat. Commun.* **13**, 6713 (2022).
20. Qin, C. *et al.* Stable room-temperature continuous-wave lasing in quasi-2D perovskite films. *Nature* **585**, 53–57 (2020).
21. Han, Y. *et al.* Living growth of ultra-bright 2D perovskites with long-lived carriers. *Nat. Commun.* **17**, 8629 (2026).
22. Liang, J. *et al.* Controlled Growth of Two-Dimensional Heterostructures: In-Plane Epitaxy or Vertical Stack. *Acc. Mater. Res.* **3**, 999–1010 (2022).
23. Chakraborty, S. K. *et al.* Seamless in two dimensions: prospects of lateral heterostructures from integration to quantum devices. *Npj 2D Mater. Appl.* **9**, 94 (2025).
24. Zhang, S. *et al.* Mosaic lateral heterostructures in two-dimensional perovskite. *Nature* **649**, 612–620 (2026).
25. Fu, Y. *et al.* Multicolor Heterostructures of Two-Dimensional Layered Halide Perovskites that Show Interlayer Energy Transfer. *J. Am. Chem. Soc.* **140**, 15675–15683 (2018).
26. Zhang, J., Zhu, X., Wang, M. & Hu, B. Establishing charge-transfer excitons in 2D perovskite heterostructures. *Nat. Commun.* **11**, 2618 (2020).

27. Borreani, M., Saini, S. K., Schleusener, A. & Krahne, R. Direct Growth of Rectangular 2D Layered Metal-Halide Perovskite Microcrystal Photonic Cavities on Functional Substrates. *Adv. Opt. Mater.* **13**, e01276 (2025).

28. Shibuya, K., Koshimizu, M., Nishikido, F., Saito, H. & Kishimoto, S. Poly[bis(phenethyl-ammonium) [di-bromido-plumbate(II)]-di-μ-bromido]]. *Acta Crystallogr. Sect. E Struct. Rep. Online* **65**, m1323–m1324 (2009).

29. Klinger, M. More features, more tools, more CrysTBox. *J. Appl. Crystallogr.* **50**, 1226–1234 (2017).

30. Schmitz, F. *et al.* Large Cation Engineering in Two-Dimensional Silver–Bismuth Bromide Double Perovskites. *Chem. Mater.* **33**, 4688–4700 (2021).

31. Connor, B. A., Leppert, L., Smith, M. D., Neaton, J. B. & Karunadasa, H. I. Layered Halide Double Perovskites: Dimensional Reduction of $Cs_2AgBiBr_6$. *J. Am. Chem. Soc.* **140**, 5235–5240 (2018).

32. Li, Z. *et al.* Core@Shell AgBr@$CsPbBr_3$ Nanocrystals as Precursors to Hollow Lead Halide Perovskite Nanocubes. *J. Am. Chem. Soc.* **147**, 23192–23201 (2025).

33. Flidr, J., Huang, Y.-C., Newton, T. A. & Hines, M. A. Extracting site-specific reaction rates from steady state surface morphologies: Kinetic Monte Carlo simulations of aqueous Si(111) etching. *J. Chem. Phys.* **108**, 5542–5553 (1998).

34. Meakin, P. & Rosso, K. M. Simple kinetic Monte Carlo models for dissolution pitting induced by crystal defects. *J. Chem. Phys.* **129**, 204106 (2008).

## Supporting Information

### Experimental Section

### Chemicals

Organic solvents including anhydrous chlorobenzene, dimethyl formamide, acetonitrile, ethyl acetate, diethyl ether, and inorganic salts including lead bromide ($PbBr_2$), lead iodide ($PbI_2$), and tin iodide ($SnI_2$) were purchased from Sigma-Aldrich. *n*-Butylammonium bromide (BA·HBr), phenethylammonium bromide (PEA·HBr), phenethylammonium iodide (PEA·HI), and n-octylammonium bromide (OA·HBr) were purchased from Greatcell Solar. Methylammonium bromide (MA.HBr) and methylammonium iodide (MA.HI) were purchased from Sigma-Aldrich. Hydroiodic acid (HI, 57 wt% in $H_2O$), hydrobromic acid (HBr, 48 wt% in $H_2O$) and hypophosphorous acid ($H_3PO_2$, 50 wt% in $H_2O$) were purchased from Sigma-Aldrich. All the above chemicals were used as received.

0.1 mm thick PDMS films and 300 nm dry thermal oxide wafers were purchased from Tera Advance.

### Synthesis of Layered Halide Perovskite microcrystals

The layered halide perovskites were synthesized by adopting the synthesis strategy reported by Song Jin and co-workers.[1]

For $BA_2PbBr_4$, $PEA_2PbBr_4$, $OA_2PbBr_4$, and $PEA_2PbI_4$ microcrystals were synthesized by dissolving the corresponding organic spacer ammonium salts (0.1M) and 0.1 M $PbX_2$ (X = Br or I) in 1mL of HBr in case of Br-based and 0.9mL HI with 0.1 mL hypophosphorous acid in case of I-based at 130°C. The solutions were cooled to approx. 50 for $PEA_2PbI_4$ and $PEA_2PbBr_4$ and to 30 for $BA_2PbBr_4$ and $OA_2PbBr_4$. Then 3µL of each solution was drop-casted on to a clean glass slide. Within 2-3 seconds microcrystals were picked up with PDMS film, dried on a Whatman paper, and then transferred on to clean oxide wafer pieces.

Higher n methylammonium (MA) based layered halide perovskites $BA_2MAPb_2Br_7$ (n = 2), $BA_2MA_2Pb_3Br_{10}$ (n = 3), and $BA_2MA_2Pb_3I_{10}$ (n = 3), microcrystals were synthesized following the exact reported recipe.

**Etching of Microcrystals**

All the microcrystals were etched by submerging the oxide wafer with microcrystals in 500 µL solution containing acetonitrile and chlorobenzene with acetonitrile varying from 100 µL up to 250 µL at various reaction temperatures varying from 0°C up to 50°C.

We also used ethyl acetate and diethyl ether as anti-solvents in place of chlorobenzene.

**Endo-epitaxial Heterostructure Synthesis**

The endo-epitaxial heterostructures were grown using a slightly modified recipe reported by Letian Dou and co-workers.[2]

Stock solution for $BA_2PbI_4$, $PEA_2PbI_4$, and $Sn_2PbI_4$ were prepared by dissolving 10 micromoles of $PbI_2$ or $SnI_2$ and 20 micromoles of BA.HI or PEA.HI, correspondingly, in 1 mL dimethyl formamide and 1 mL chlorobenzene solution.

**Endo-epitaxial $PEA_2PbI_4$-$PEA_2PbBr_4$ Heterostructures**

The $PEA_2PbI_4$-$PEA_2PbBr_4$ heterostructures were synthesized by drop casting 5 µL of a 250 times diluted solution at 35°C on top of the etched microcrystals a hot plate. The diluted precursor solution was prepared by diluting the $PEA_2PbI_4$ stock solution approx. 500 times by adding a 5 µL of stock solution to 2350 µL chlorobenzene and 150 µL acetonitrile solution.

For increasing the size of newly grown $PEA_2PbI_4$ domain, we added the diluted solution in cycles to obtain desired size.

**Endo-epitaxial $BA_2PbI_4$-$BA_2PbBr_4$ Heterostructures**

The $BA_2PbI_4$-$BA_2PbBr_4$ heterostructures were synthesized by drop casting 5 µL of a 250 times diluted solution at 35°C on top of the etched microcrystals a hot plate. The diluted precursor solution was prepared by diluting the $BA_2PbI_4$ stock solution approx. 500 times by adding a 5 µL of stock solution to 2350 µL chlorobenzene and 150 µL acetonitrile solution.

**Endo-epitaxial $PEA_2SnI_4$-$PEA_2PbBr_4$ Heterostructures**

The $PEA_2SnI_4$-$PEA_2PbBr_4$ heterostructures were synthesized by drop casting 5 µL of a 250 times diluted solution at 35°C on top of the etched microcrystals a hot plate. The diluted precursor solution was prepared by diluting the $PEA_2SnI_4$ stock solution approx. 500 times by adding a 5 µL of stock solution to 2350 µL chlorobenzene and 150 µL acetonitrile solution.

For TEM analysis, the $PEA_2PbBr_4$ microcrystals were synthesized directly on top of Cu grids with carbon support.

The stock solution of $PEA_2PbBr_4$ was prepared by dissolving 10 micromoles of $PbBr_2$ and 20 micromoles of PEA.HBr in 1 mL dimethyl formamide and 1 mL chlorobenzene solution. The 10 μL of the stock solution was diluted by adding to 800 μL chlorobenzene and 200 μL acetonitrile solution. Then 5 μL of the diluted solution was drop casted on top of the TEM grid at 75°C.

**Optical Characterization**

The bright-field optical images were collected using an optical profilometer zeta microscope. The optical fluorescence images were collected using Nikon A1 confocal microscope using 405 nm laser as excitation source. The PL spectra from individual microcrystals were collected using a home built optical setup.[3]

**Structural Characterization**

The X-ray diffraction patterns were collected using Malvern Panalytical Empyrean diffractogram with Cu Kα source. The transmission electron microscope images and corresponding selected area electron diffraction (SAED) patterns were collected using JEOL JEM-1400Plus – Analytical 120 kV TEM. The SAED patterns were indexed using CrystBox.[4] AFM Park was used for collecting height and surface profiles.

**Cavity Model Simulation**

The growth of the etched cavities was simulated with a three-dimensional, rejection-free kinetic Monte Carlo (kMC) model based on the n-fold-way algorithm of Bortz, Kalos and Lebowitz.[5,6] The film is divided into volume elements (sites) that are removed one at a time, at a rate that depends on the number of neighboring sites that have already been removed. The sites are coarse-grained volume elements, not atoms. The aim of the model is to reproduce the measured cavity geometry (depth, length, width and the shape of the cross-section) with a small number of effective rates. It does not describe chemistry or the transport of the etchant, and the fitted rates should not be read as physical rate constants. The model is phenomenological: it reproduces the measured cavity geometry with a small set of effective rates, not with a derived dissolution mechanism.

**Lattice**

Based on our experiment, we consider a 75 nm thick microcrystal as a cubic lattice of sites with x along [010] (cavity width), y along [100] (cavity length) and z along the film depth. At the coarsest resolution a site is 50 nm in the film plane. The depth is divided into 30 slabs, so a slab is 2.5 nm thick. A resolution factor fl refines both: fl = 3 gives 16.7 nm pixels and 0.83 nm slabs. The substrate stops the vertical etch at 75 nm. A site can dissolve only if it is exposed, so the void in each column stays connected to the surface and there is no undercutting. The simulation box is 2.0 µm wide and (3.0 + 1.5·d) µm long, where d is the distance between the two cavity seeds. A run stops early if a cavity reaches the box edge, 3.2 µm in width or 4.4 µm in length.

**Removal rates and time stepping**

Etching starts from a small seed (2 × 4 sites, i.e. 100 × 200 nm at fl = 1) in the top slab. Each exposed site dissolves with a rate that depends on its void neighbors, and the next event is chosen from the rate-weighted list of site classes. Two processes remove a site. Lateral removal advances the cavity edges. A site in a layer at depth z with nx and ny removed neighbors along x and y is removed at the rate

$$r_lat = g(z) \cdot \kappa \cdot ( n_x + r_y \cdot n_y )$$

where $r_y = 1.8$ is the ratio of the lateral rates along y ([100], the long axis) and x ([010], the short axis). The factor $\kappa = 4$ applies if the site has removed neighbors along both x and y (a kink in a step edge). It makes an edge that has started to move clear as a whole row, so the openings stay close to rectangular, with straight edges and sharp corners. The factor g(z) describes the decrease of the lateral rate with depth:

$$g(z) = \max( \exp(-z / z_L) , g_{deep} )$$

with $z_L = 15$ nm and $g_{deep} = 0.3$. It equals 1 at the surface, falls to 0.37 at 15 nm and reaches its floor of 0.3 at a depth of about 18 nm. The upper walls of the cavity therefore retreat faster than the lower ones and form a conical mouth, below which the walls are steeper.

Vertical removal deepens the cavity. A site below the top layer is removed at the rate

$$k_z(\tau) = r_z \cdot \tau^q , \quad q = 1$$

only if it lies inside a central channel. The vertical rate increases with time because the measured depth rate does not decrease during the etching. A site outside the channel can be reached only laterally.

Sites with the same layer, the same ($n_x$, $n_y$) and the same channel membership have the same rate and form one class. With $N_c$ sites of rate $r_c$ in class c, the total rate is $R = \Sigma N_c r_c$. In each step the simulation time advances by $\Delta\tau = -\ln(\xi)/R$, where $\xi$ is a uniform random number on (0, 1]. A class is chosen with probability $N_c r_c/R$, and one of its sites is chosen at random and removed. The classes of the removed site and of its neighbors are then updated. The lateral rate along x at the surface is the unit of simulation time.

**Central Channel**

The measured cross-sections show a wide, conical opening that goes over into a narrow shaft. The model describes the opening by the depth dependence of the lateral rate (Section 2.3) and the shaft by restricting the vertical removal to a central channel. The channel is a rectangle around the seed with a half-width hx along x and the same half-width along y. It widens with time, so that the shaft is narrow at first and becomes wider as the etching proceeds:

$$h_x(\tau) = h_0 + (h_1 - h_0) \cdot \min( c_max , (\tau/\tau_c)\hat{}s )$$

with h0 = 50 nm, h1 = 200 nm, s = 4 and cmax = 2, so that the half-width saturates at 350 nm. The time τc is chosen such that hx = h1 at an etching time of 11 min. The channel is a model device for the shaft observed in the AFM profiles and not a derived mechanism.

**Mapping simulation time to etching time**

The etching time t is related to the simulation time τ by a sublinear power law,

$$\tau = u \cdot t_ref \cdot ( t / t_ref )\hat{}p , \quad p = 0.4 , \quad t_ref = 5 \text{ min}$$

which gives a fast opening in the first minutes and slower growth later. The factor u is a fit parameter. Snapshots of the cavity were taken at the values of τ that correspond to etching times of 2, 4, 6, 11 and 30 min.

**Second Cavity and Merging**

At 30 minutes of etching, we observed the coalescence of two adjacent cavities (see discussion of **Figure 5**), resulting in the formation of a single larger cavity. A second cavity was therefore simulated as an independent run of the same model. It starts at t2 = 4 min from a seed at a distance d = 0.3 μm from the first seed along y. Its clock is slower than that of the first cavity: at etching time t it is in the state that cavity 1 has at the time f2 (t − t2), with f2 = 0.4. The simulated sample is the union of the two removed volumes. In each layer the two cavities are count as merged when their extents along y touch or overlap. From then on the length, width and depth are those of the merged cavity, and before that they are those of cavity 1. The cavities do not influence each other before they touch, and after merging each keeps growing by its own rules (union approximation).

**Calibration**

The model was calibrated against the AFM data. The vertical rate prefactor rz and the time scale u were fitted by a grid search with 10 independent runs per grid point. The objective is $\chi^2 = \chi^2_{size} + \chi^2_{shape}$, with

$$\chi^2_size = \Sigma_i \left[ ((d_i - \hat{d}_i)/\sigma_{d,i})^2 + ((l_i - \hat{l}_i)/(0.08\, \hat{l}_i))^2 + ((w_i - \hat{w}_i)/(0.08\, \hat{w}_i))^2 \right]$$

where d, l and w are the depth, length and width at the five etching times i, hats denote measured values, and $\sigma_{d,i} = \max(0.08\, \hat{d}_i, 2\text{ nm})$. At 30 min the length is compared with that of the merged cavity (1.91 µm). The shape term compares the extent λ of the cavity along y at 10, 20, …, 60 nm below the surface with the measured cross-section of hole 1:

$$\chi^2_shape = \Sigma_{i,j} ((\lambda_{ij} - \hat{\lambda}_{ij})/(0.15\, \hat{\lambda}_{ij} + 0.05\ \mu m))^2$$

A level that exists in only one of simulation and measurement adds a fixed penalty of 4. The tolerances are weights of the fit and not measurement uncertainties. All other parameters (ry, κ, zL, gdeep, the channel parameters, the seed, t2, d and f2) were set by hand to reproduce the qualitative features of the measured cross-sections and were not optimized. The best grid point, rz = 15 (in units of the coarsest lattice) and u = 0.23, was rerun 20 times. The results are given as the mean ± standard deviation over these runs, and the profiles and maps shown in the figures are from one representative run.

**Implementation and limitations**

The model is implemented in MATLAB. Each run uses its own random number seed, and runs are independent. The depth of a cavity is the number of removed layers in its deepest column multiplied by the layer thickness. Length and width are the extents of the removed region in the top layer along y and x, and the extent at a depth is the extent along y of the removed region in that layer.

The channel and the depth dependence of the lateral rate describe the geometry of the cavities but do not follow from a mechanism. The model has no etchant transport, no crystallographic detail beyond the ratio ry, and no undercutting, and most of its parameters were set by hand. The two cavities are independent until they touch, and the results depend somewhat on the lattice resolution.

**Table S1: Simulation Parameters**

| Parameter | Meaning | Value | Determination |
|---|---|---|---|
| Lattice | Lateral site size, layer thickness, film thickness | 16.7 nm, 0.83 nm, 75 nm | Numerical, film from experiment |
| $r$y | Ratio of the lateral rates along [100] (y) and [010] (x) | 1.8 | Set by hand |
| κ | Kink enhancement of the lateral rate | 4 | Set by hand |
| $z$L, $g$deep | Depth dependence of the lateral rate | 15 nm, 0.3 | Set by hand |

| Parameter | Meaning | Value | Determination |
| --- | --- | --- | --- |
| *r*z, *q* | Prefactor and exponent of the vertical rate | 15 (coarsest-lattice units), 1 | *r*z fitted, *q* fixed |
| *u*, *p*, *t*ref | Time mapping | 0.23, 0.4, 5 min | *u* fitted, others set by hand |
| *h*0, *h*1, *s*, *c*max | Central channel | 50 nm, 200 nm, 4, 2 | Set by hand |
| Seed | Initial removed region | 100 × 200 nm | Set by hand |
| *t*2, *d*, *f*2 | Emergence time, seed distance and relative speed of cavity 2 | 4 min, 0.3 μm, 0.4 | Set by hand |
| Runs | Per grid point, at the best point | 10, 20 | Numerical |

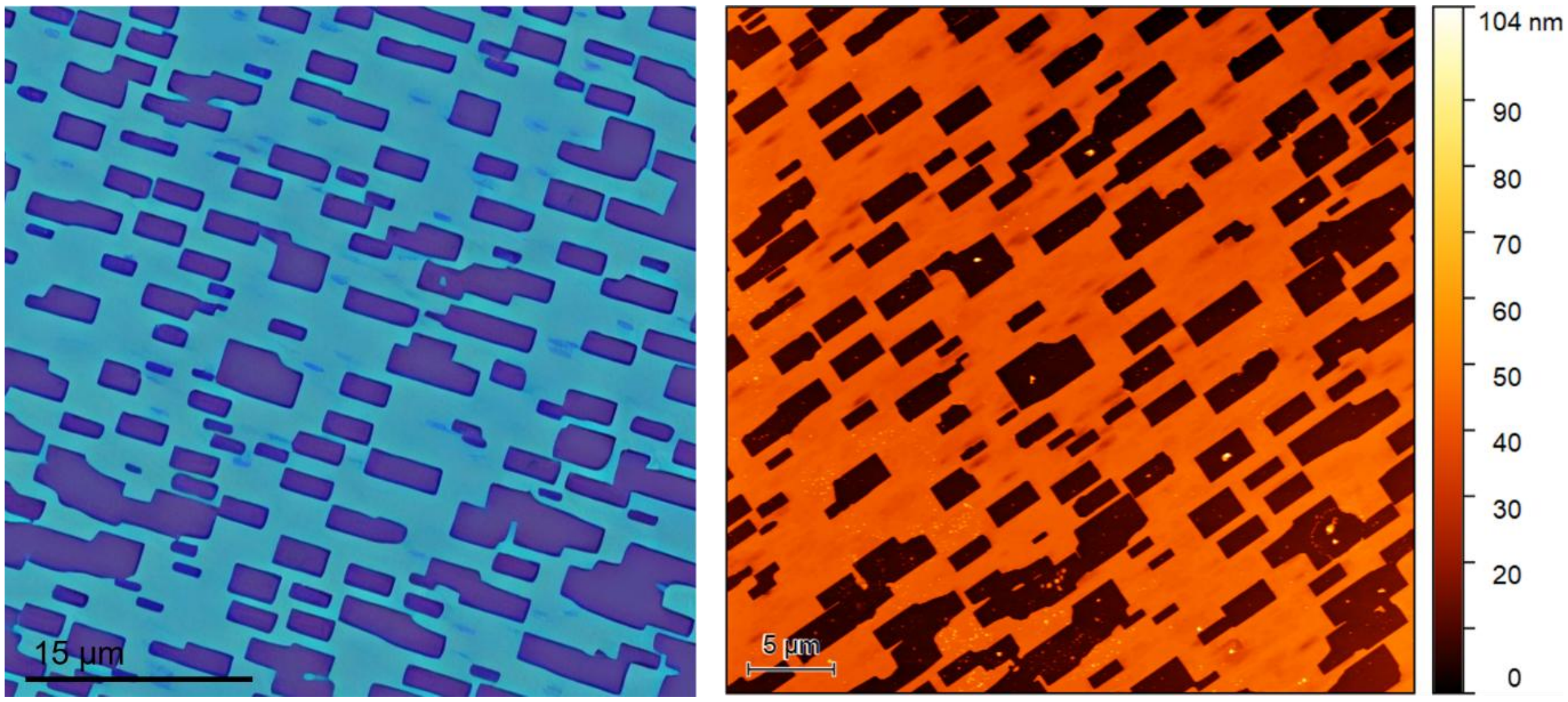

**Figure S1 | Optical and AFM characterization of etched $PEA_2PbBr_4$ microcrystals.** Optical BF image of etched $PEA_2PbBr_4$ microcrystal and the corresponding AFM image.

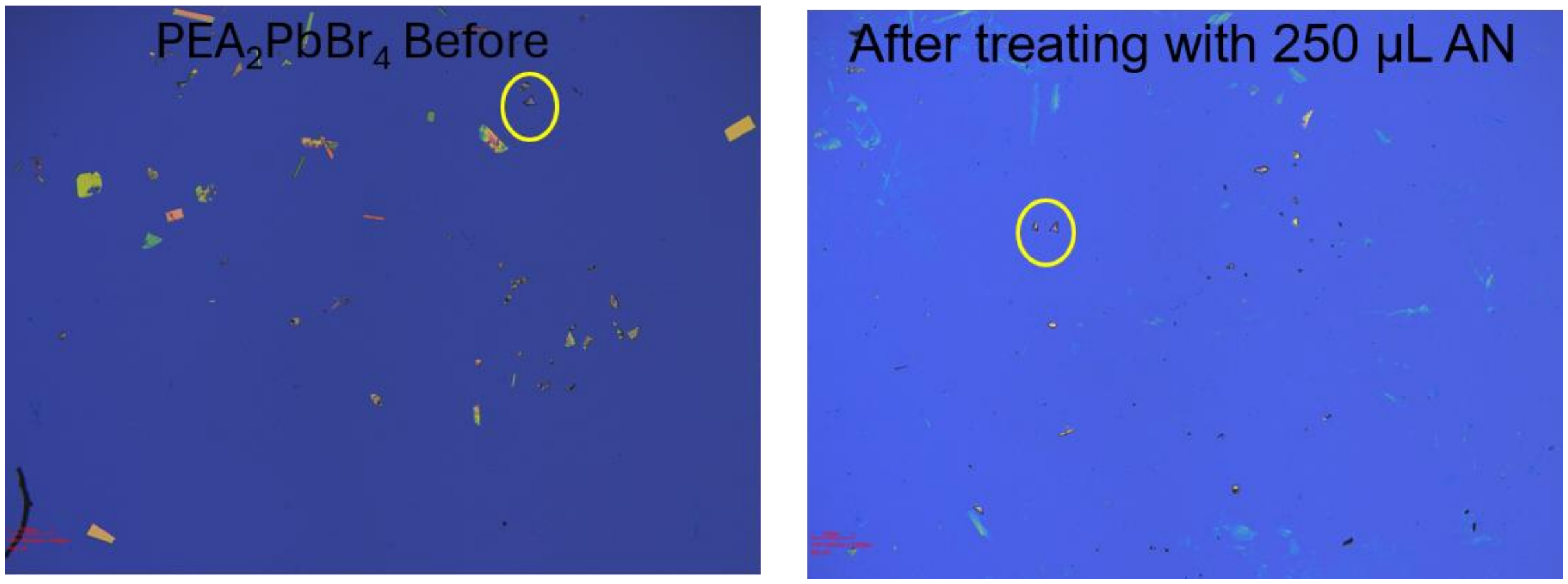

**Figure S2 | Dissolution of $PEA_2PbBr_4$ microcrystals at high acetonitrile content.** $PEA_2PbBr_4$ microcrystals treated with 250 μL acetonitrile (AN) and 250 μL chlorobenzene solution at room temperature. Yellow circle encircles the 2 surviving microcrystals.

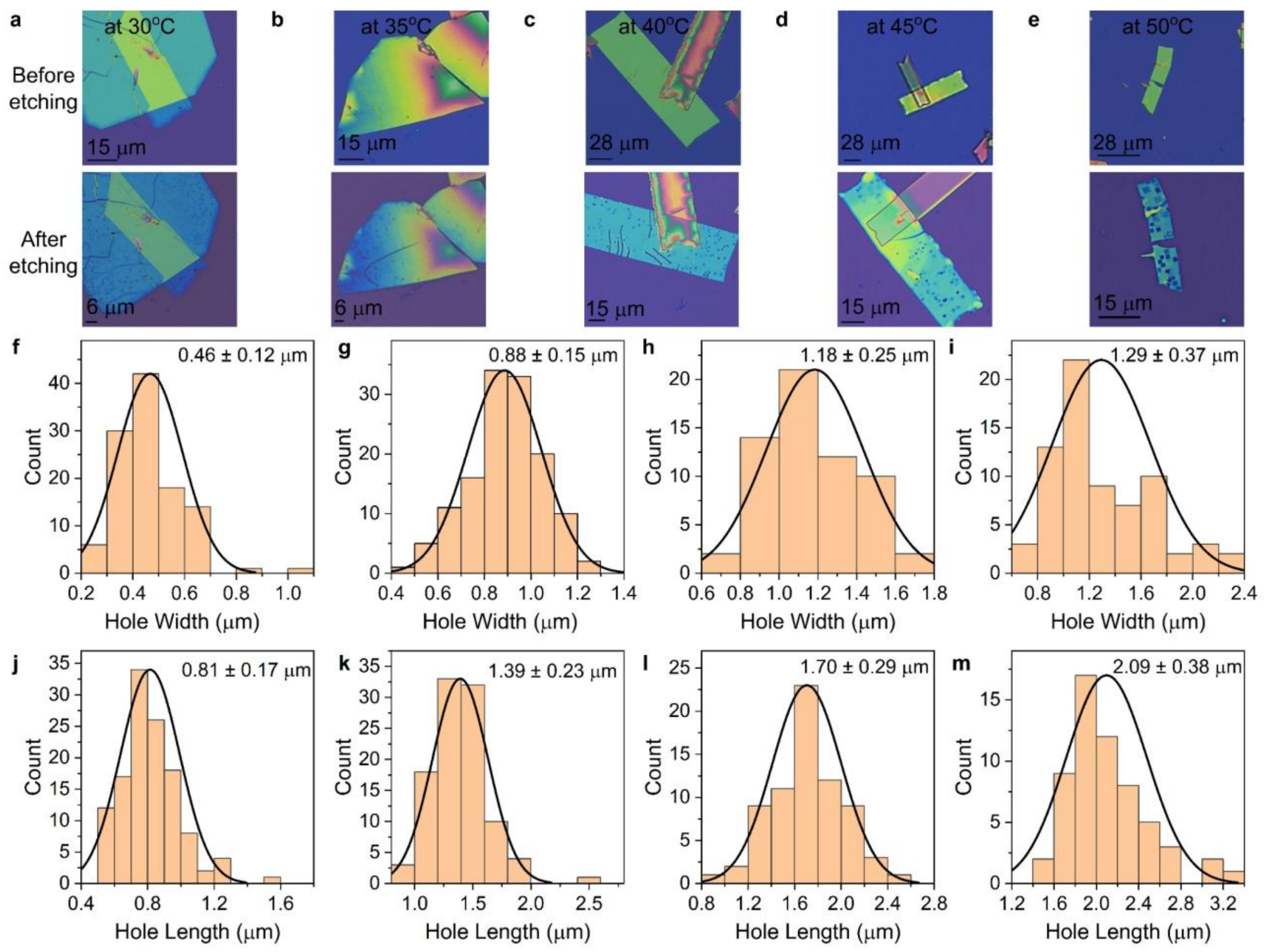

**Figure S3 | Temperature-dependent etching of $PEA_2PbBr_4$ microcrystals. a-e** Optical BF images of pre-etched and after etching $PEA_2PbBr_4$ microcrystals at 30°C, 35°C, 40°C, 45°C, and 50°C, respectively. **f-i** Hole width distribution and **j-m** hole length distribution corresponding to etching done at 35°C, 40°C, 45°C, and 50°C, respectively. At 30°C the etched holes were too small within 1 min of etching duration for all the samples.

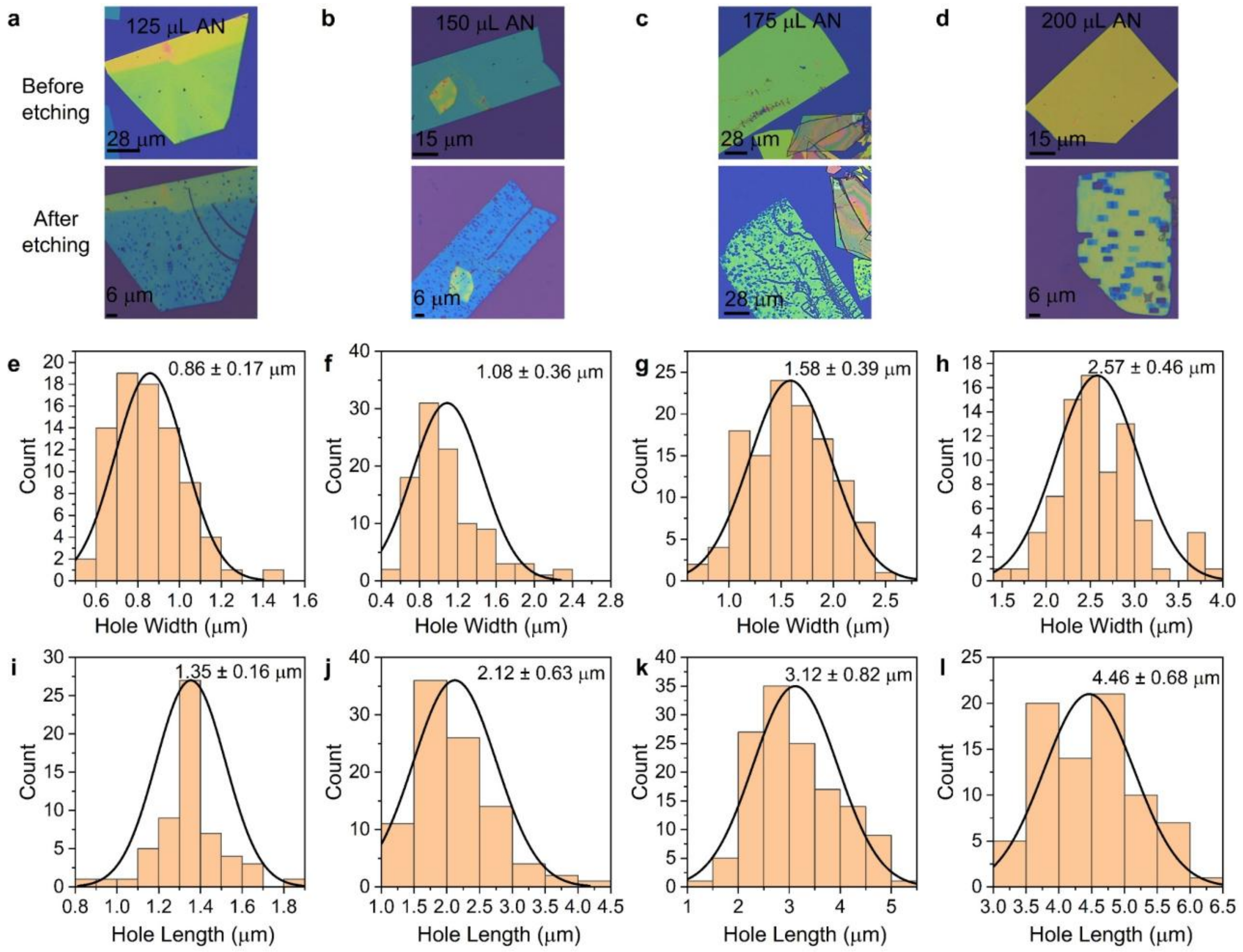

**Figure S4 | Effect of acetonitrile volume on hole formation in $PEA_2PbBr_4$ microcrystals. a-d** Optical BF images of pre-etched and after etching $PEA_2PbBr_4$ microcrystals at room temperature after 1 min etching with 125 μL, 150 μL, 175 μL, and 200 μL volume of acetonitrile in a total solution volume of 500 μL, respectively. **e-h** Hole width distribution and **i-l** hole length distribution plots corresponding to etching with 125 μL, 150 μL, 175 μL, and 200 μL volume of acetonitrile, respectively.

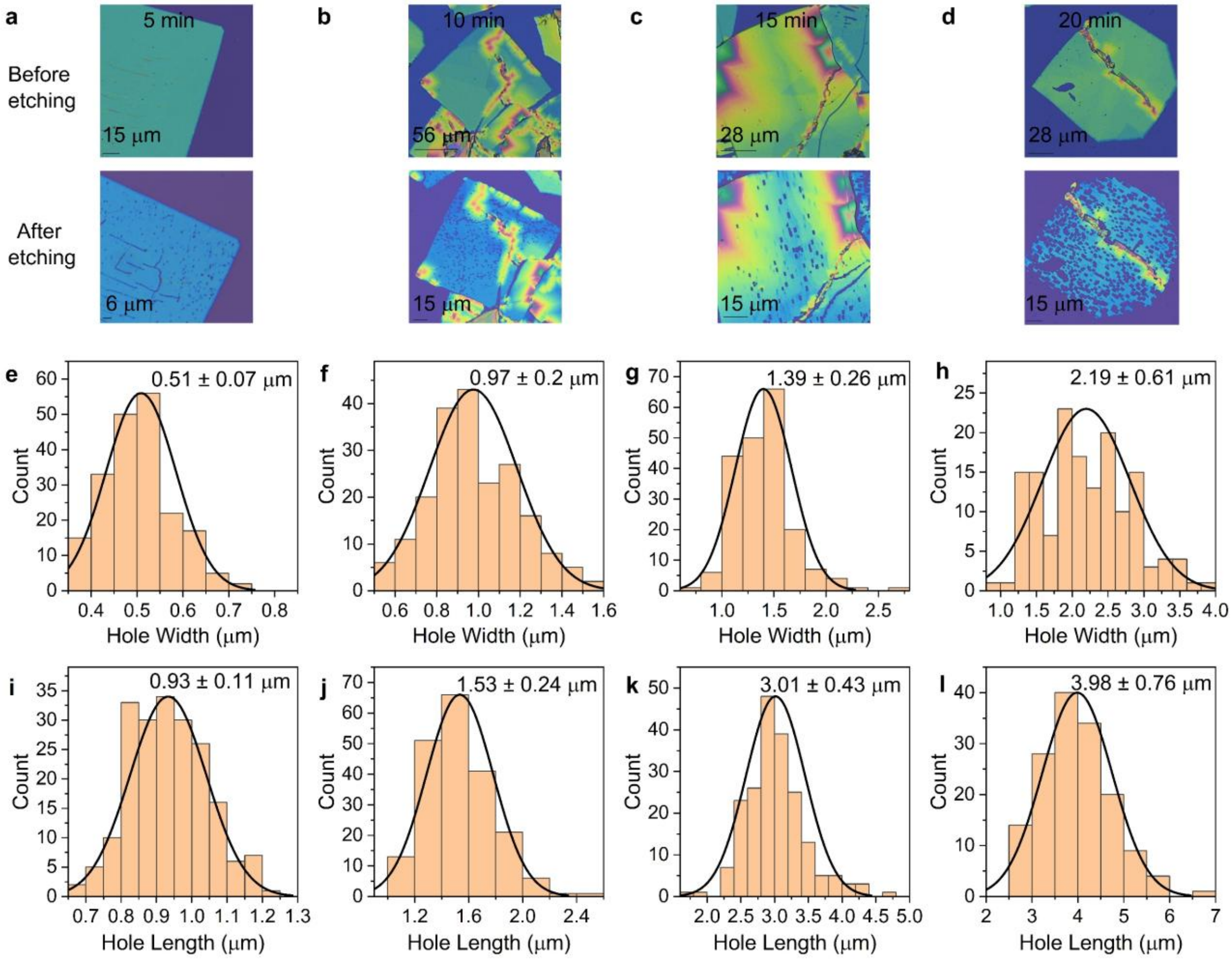

**Figure S5 | Time-dependent evolution of etched holes in $PEA_2PbBr_4$ microcrystals. a-d** Optical BF images of pre-etched and after etching $PEA_2PbBr_4$ microcrystals at room temperature for 5 min, 10 min, 15 min, and 20 min, respectively. **e-h** Hole width distribution and **i-l** hole length distribution plots corresponding to etching duration of 5 min, 10 min, 15 min, and 20 min, respectively.

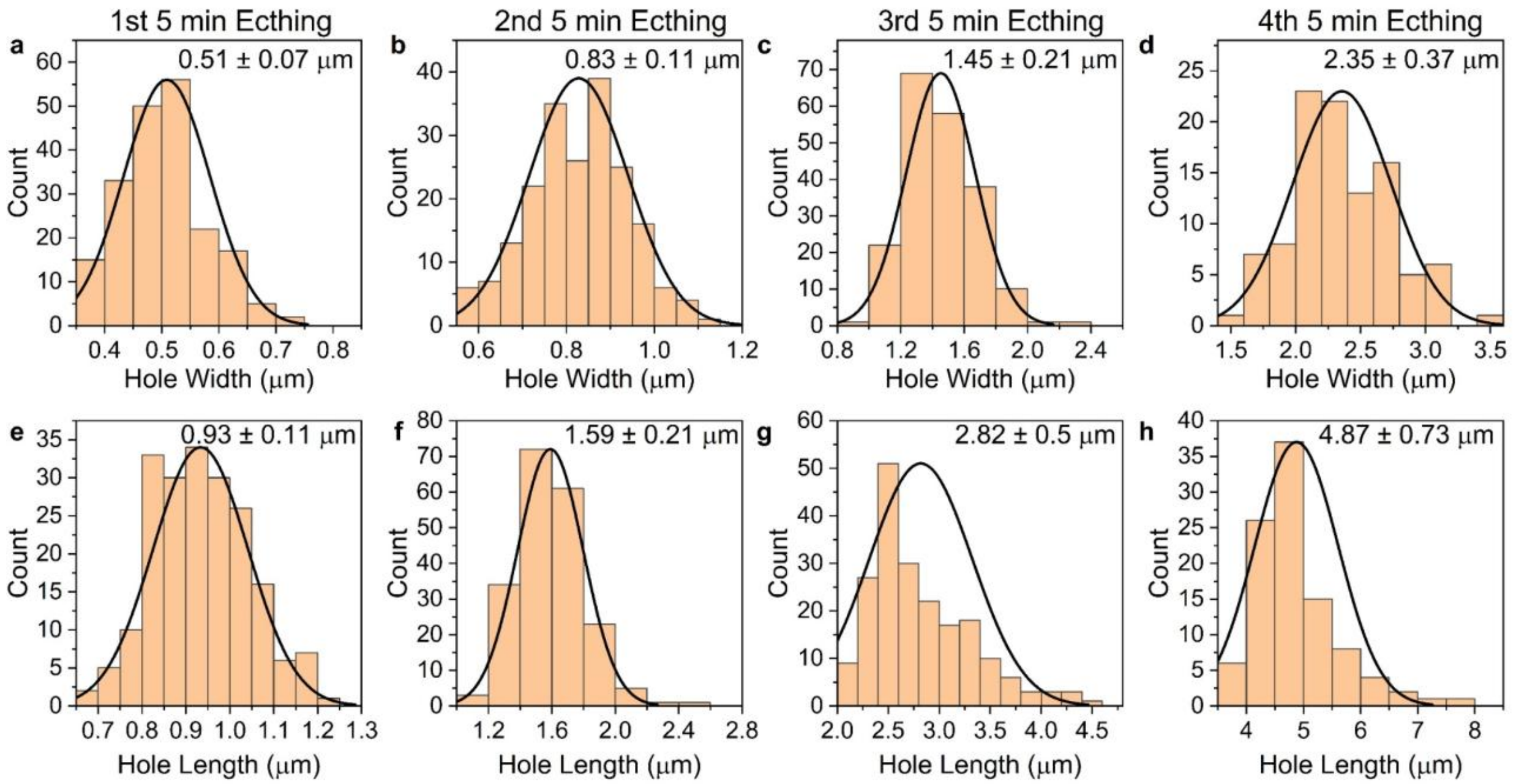

**Figure S6 | Hole-size evolution during successive etching cycles. a-d** Hole width distribution and **e-h** hole length distribution plots corresponding to etching duration of 5 min for each cycle, respectively.

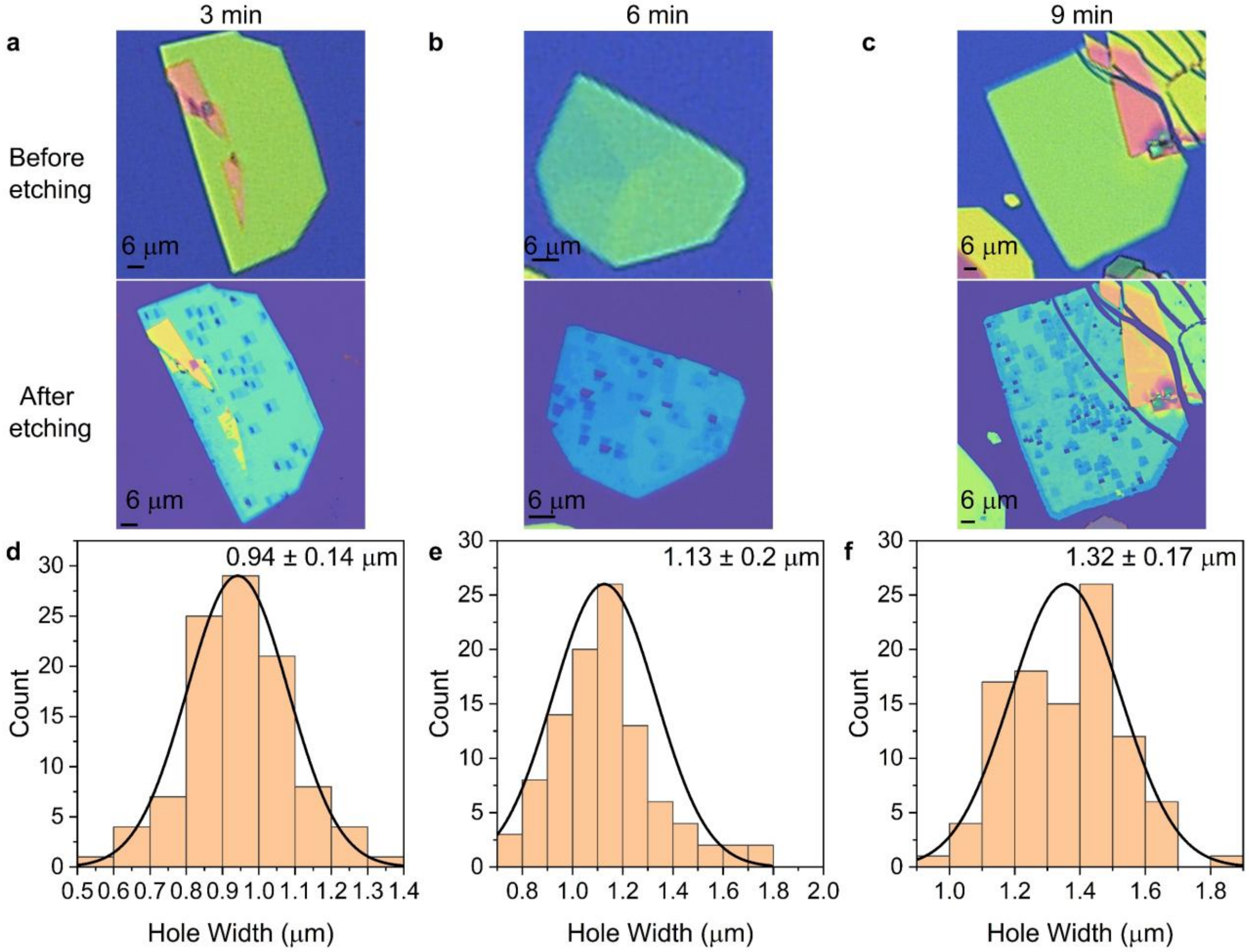

**Figure S7 | Time-dependent hole formation in $PEA_2PbBr_4$ microcrystals at 35°C. a-c** Optical BF images of pre-etched and after etching $PEA_2PbBr_4$ microcrystals at 35°C after etching duration of 3 min, 6 min, and 9 min, respectively. **d-f** Hole width distribution plots corresponding to etching duration of 3 min, 6 min, and 9 min, respectively.

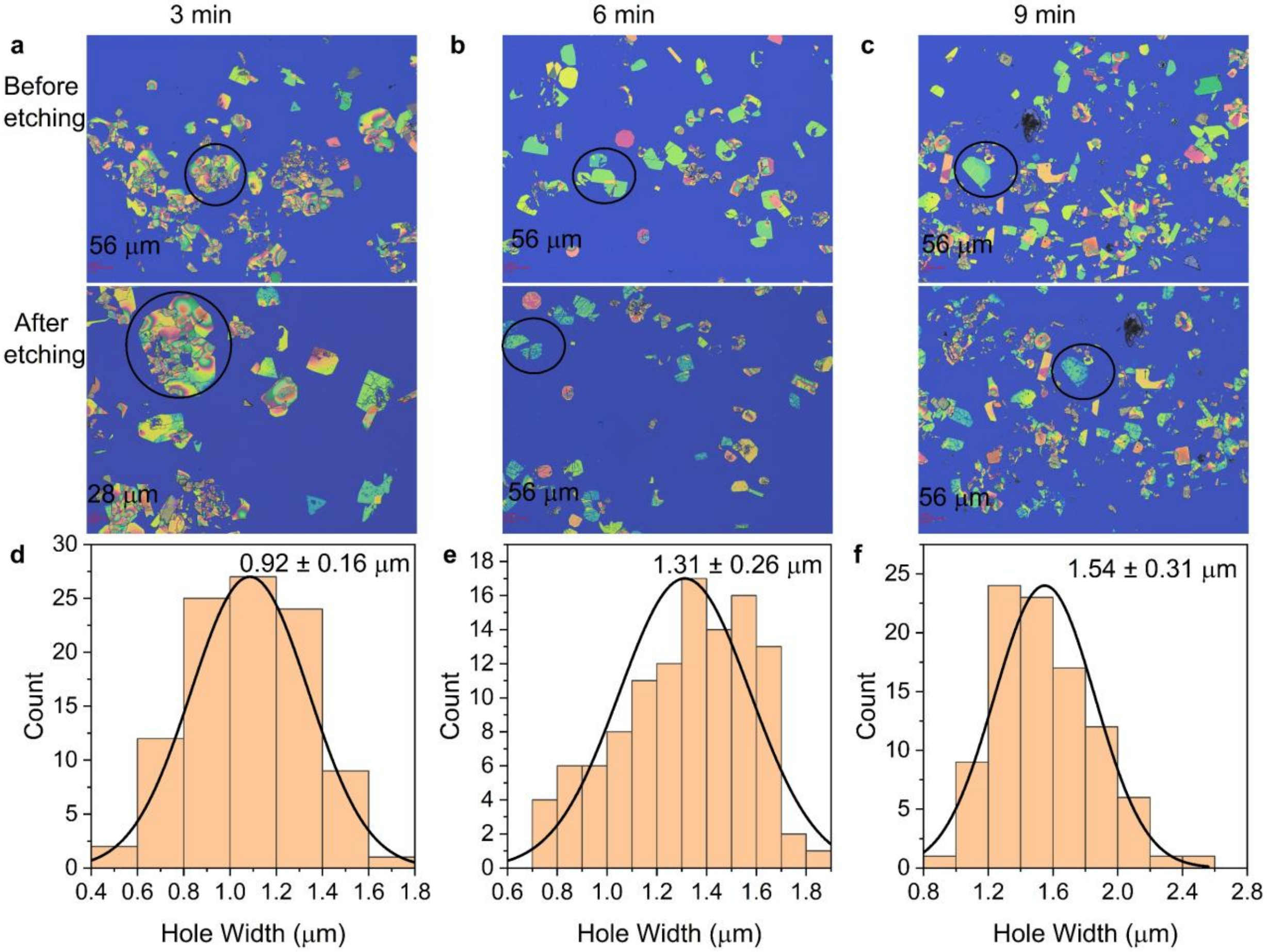

**Figure S8 | Large-area optical characterization of $PEA_2PbBr_4$ etching at 40°C. a-c** Large area optical BF images of pre-etched and after etching $PEA_2PbBr_4$ microcrystals at 40°C after etching duration of 3 min, 6 min, and 9 min, respectively. **d-f** Hole width distribution plots corresponding to etching duration of 3 min, 6 min, and 9 min, respectively. The circled microcrystals for pre-etched and etched can be used to navigate the large BF images.

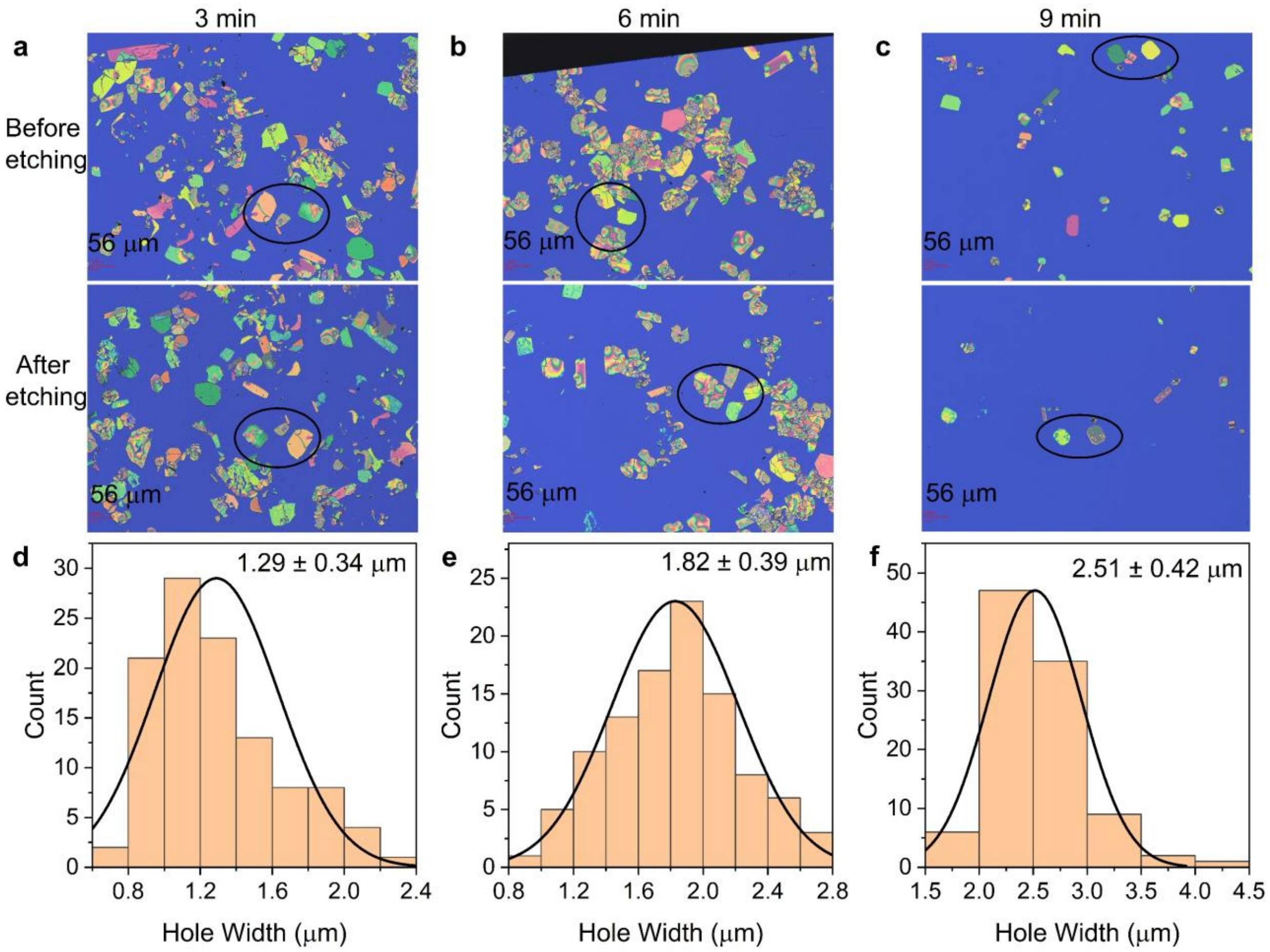

**Figure S9 | Large-area optical characterization of $PEA_2PbBr_4$ etching at 45°C. a-c** Large area optical BF images of pre-etched and after etching $PEA_2PbBr_4$ microcrystals at 45°C after etching duration of 3 min, 6 min, and 9 min, respectively. **d-f** Hole width distribution plots corresponding to etching duration of 3 min, 6 min, and 9 min, respectively. The circled microcrystals for pre-etched and etched can be used to navigate the large BF images.

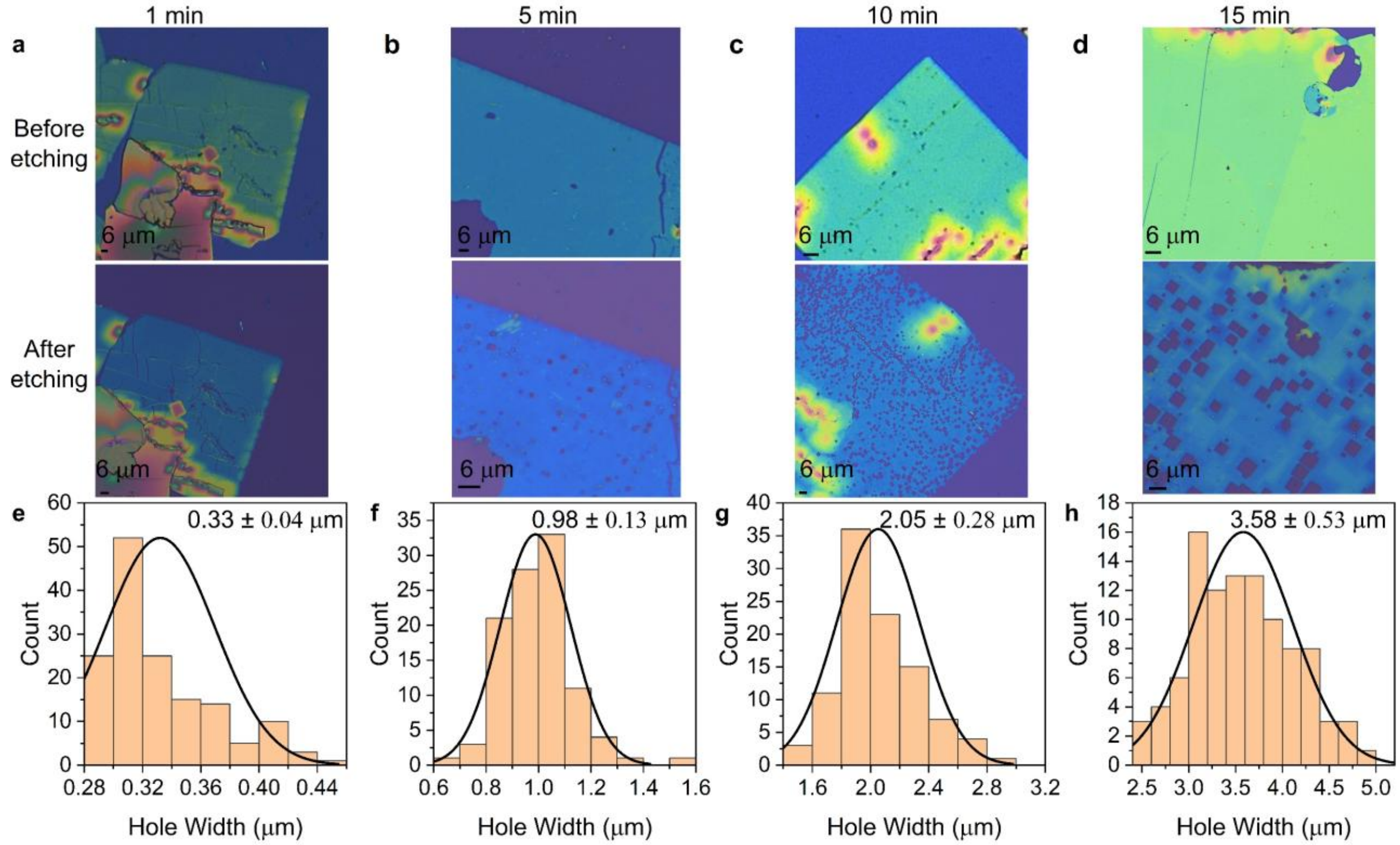

**Figure S10 | Time-dependent etching of $BA_2PbBr_4$ microcrystals at 25°C. a-d** Optical BF images of pre-etched and after etching $BA_2PbBr_4$ microcrystals at 25°C after etching duration of 1 min, 5 min, 10 min, and 15 min, respectively. **e-h** Hole size distribution plots corresponding to etching duration of 1 min, 5 min, 10 min, and 15 min, respectively.

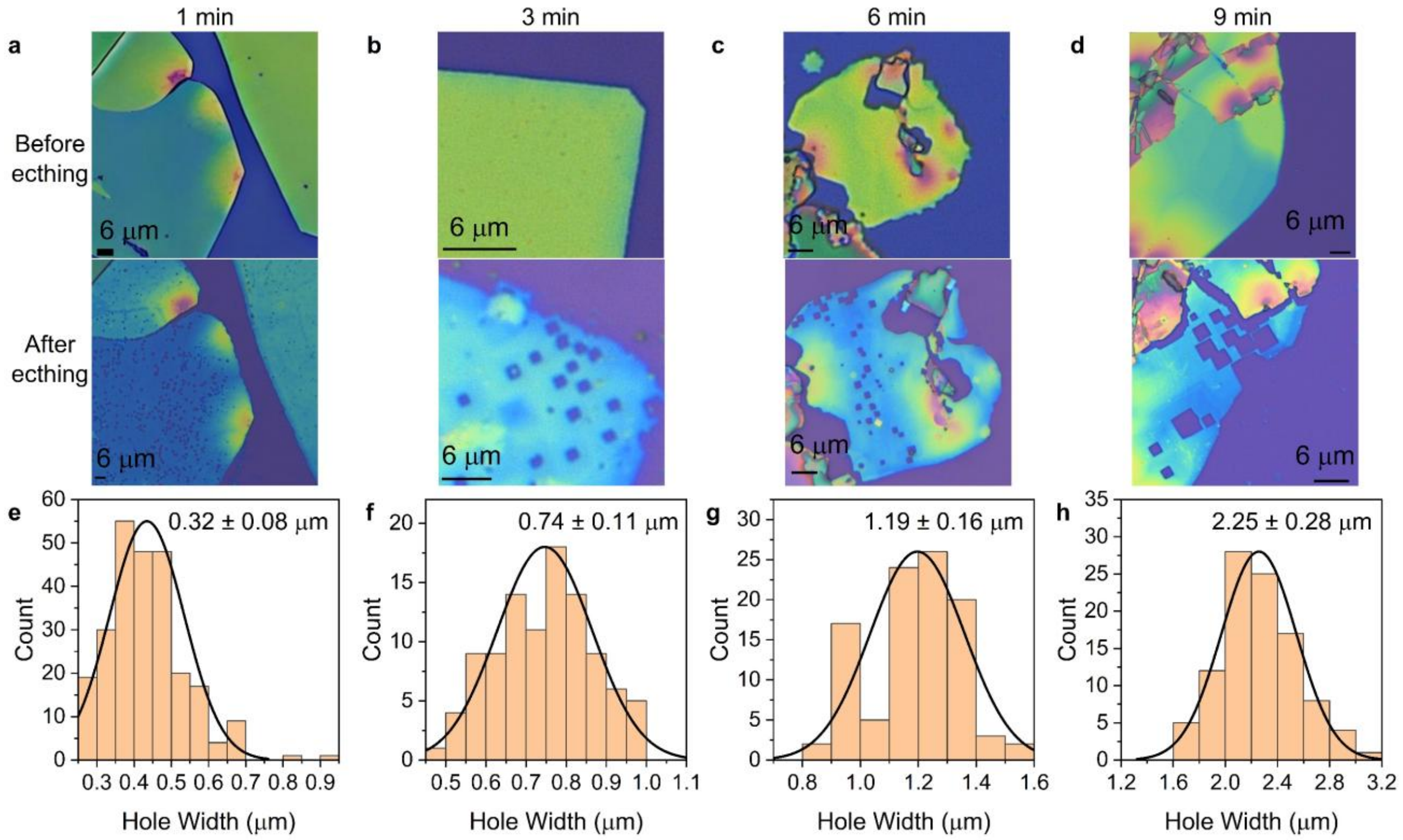

**Figure S11 | Time-dependent etching of $BA_2PbBr_4$ microcrystals at 30°C. a-d** Optical BF images of pre-etched and after etching $BA_2PbBr_4$ microcrystals at 30°C after etching duration of 1 min, 3 min, 6 min, and 9 min, respectively. **e-h** Hole size distribution plots corresponding to etching duration of 1 min, 3 min, 6 min, and 9 min, respectively.

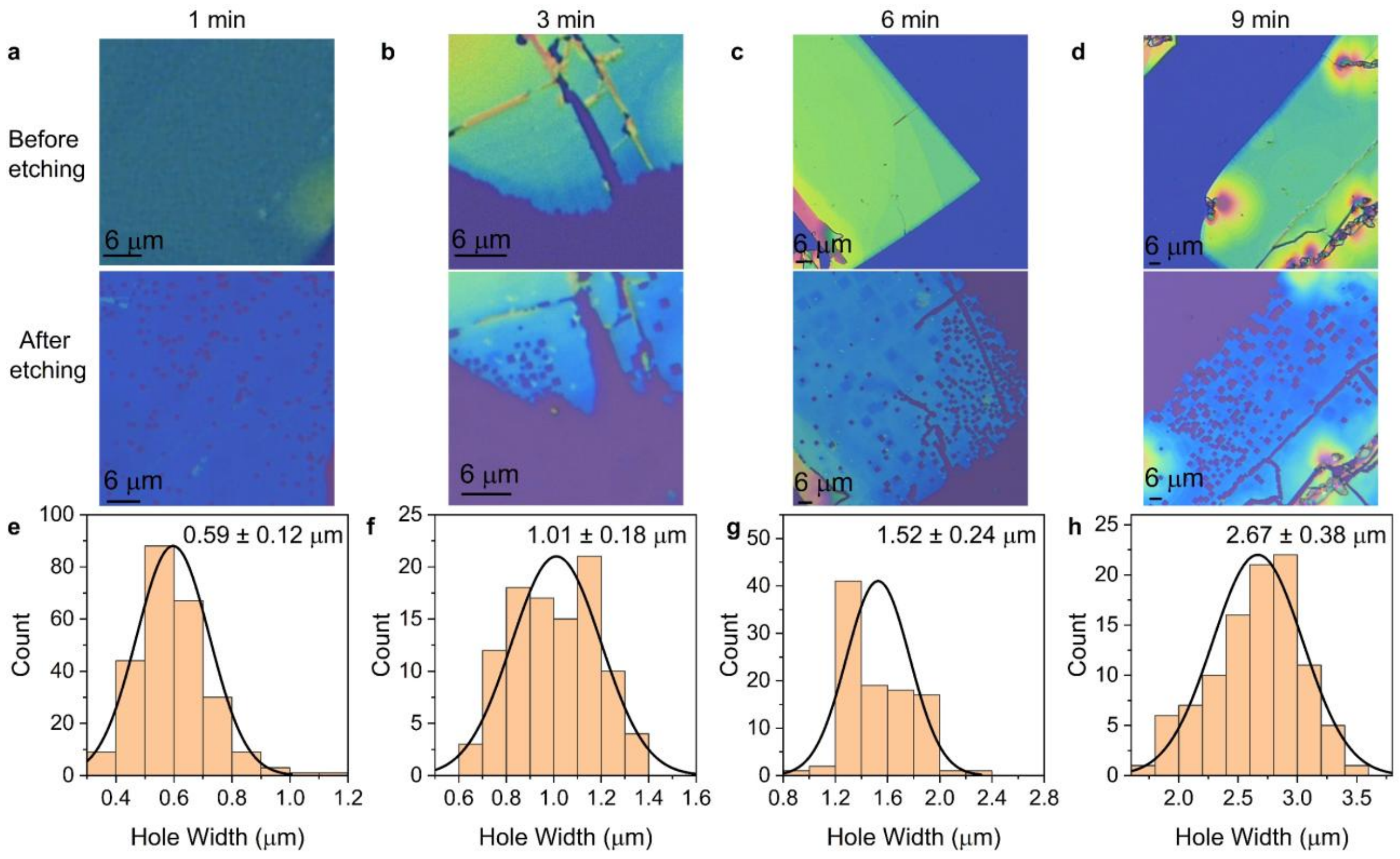

**Figure S12 | Time-dependent etching of $BA_2PbBr_4$ microcrystals at 35°C. a-d** Optical BF images of pre-etched and after etching $BA_2PbBr_4$ microcrystals at 35°C after etching duration of 1 min, 3 min, 6 min, and 9 min, respectively. **e-h** Hole size distribution plots corresponding to etching duration of 1 min, 3 min, 6 min, and 9 min, respectively.

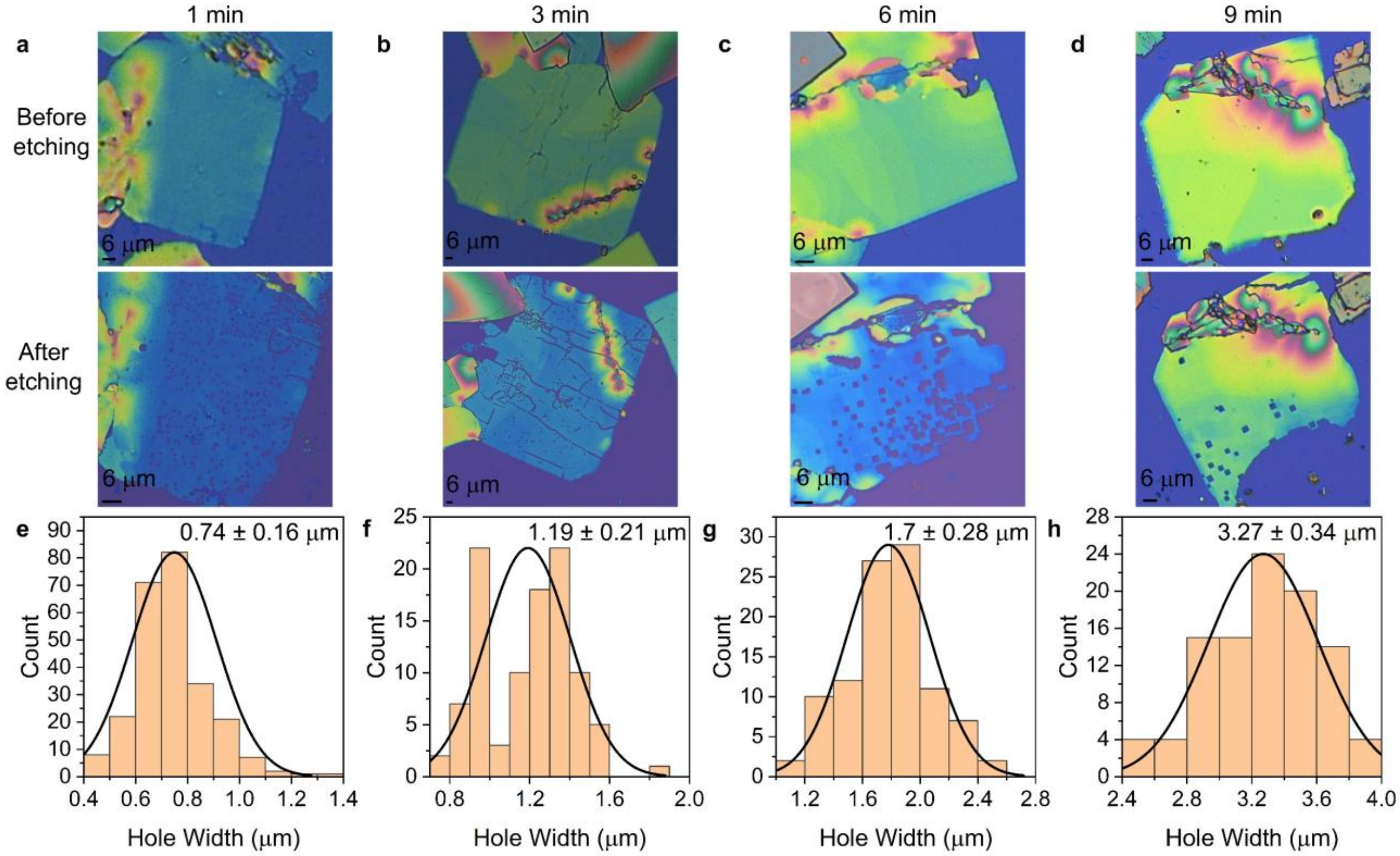

**Figure S13 | Time-dependent etching of $BA_2PbBr_4$ microcrystals at 40°C. a-d** Optical BF images of pre-etched and after etching $BA_2PbBr_4$ microcrystals at 40°C after etching duration of 1 min, 3 min, 6 min, and 9 min, respectively. **e-h** Hole size distribution plots corresponding to etching duration of 1 min, 3 min, 6 min, and 9 min, respectively.

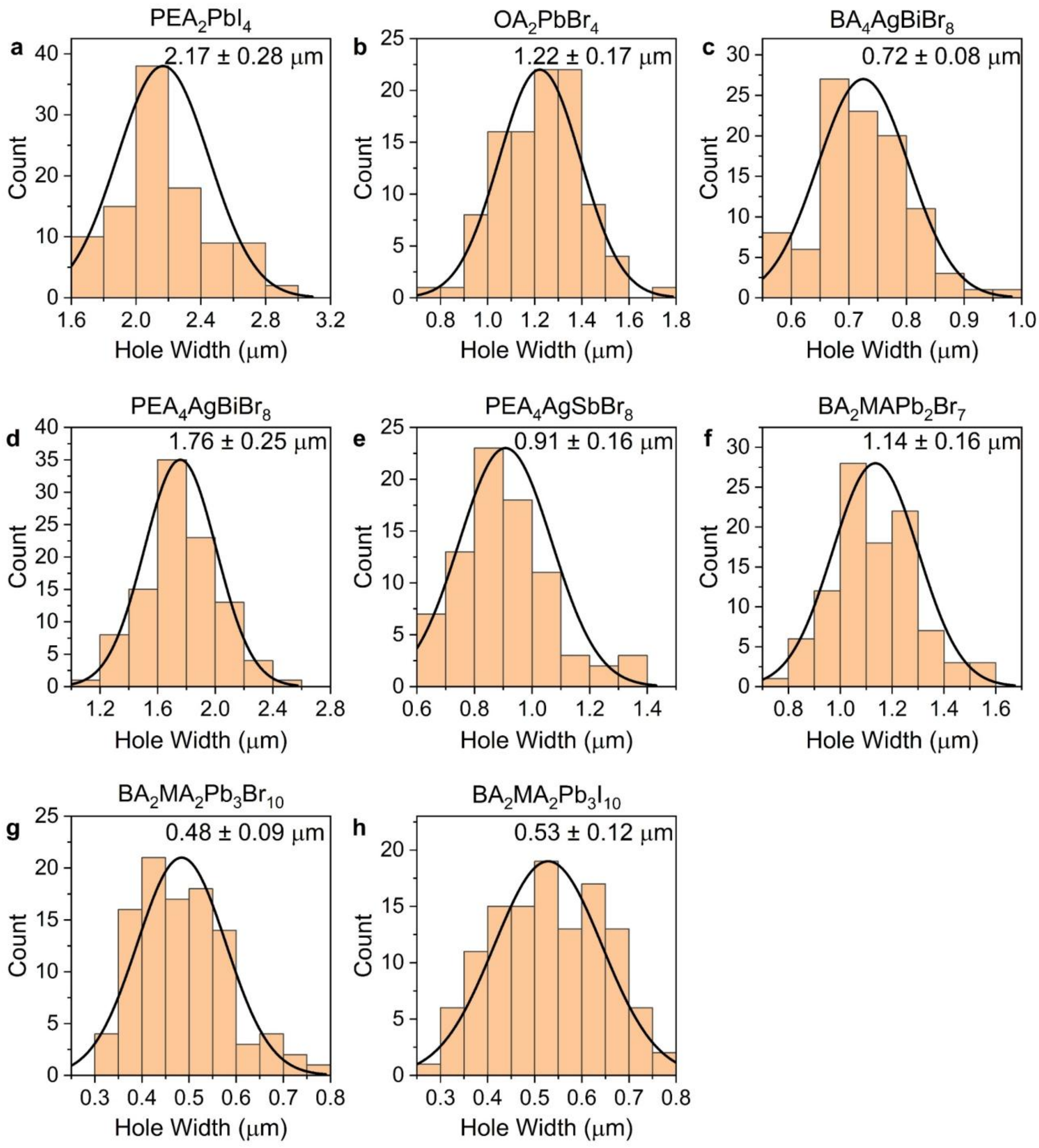

**Figure S14 | Hole size distribution plots of various etched perovskite microcrystals. a** $PEA_2PbI_4$, **b** $OA_2PbBr_4$, **c** $BA_4AgBiBr_8$, **d** $PEA_4AgBiBr_8$, **e** $PEA_4AgSbBr_8$, **f** $BA_2MAPb_2Br_7$, **g** $BA_2MA_2Pb_3Br_{10}$, and **h** $BA_2MA_2Pb_3I_{10}$.

**Table S2:** Perovskites and the etching solution and etching reaction duration corresponding to the optical images in **Figure 4**.

| Perovskite | Acetonitrile Volume (μL) | Chlorobenzene (μL) | Etching Reaction Duration (seconds) |
|---|---|---|---|
| $BA_2PbBr_4$ | 150 | 350 | 60 |
| $PEA_2PbI_4$ | 100 | 400 | 90 |
| $OA_2PbBr_4$ | 150 | 350 | 60 |
| $BA_4AgBiBr_8$ | 10 | 490 | 5 |
| $PEA_4AgBiBr_8$ | 100 | 400 | 120 |
| $PEA_4AgSbBr_8$ | 100 | 400 | 60 |
| (n = 2) $BA_2MAPb_2Br_7$ | 150 | 400 | 60 |
| (n = 3) $BA_2MA_2Pb_3Br_{10}$ | 150 | 400 | 60 |
| (n = 3) $BA_2MA_2Pb_3I_{10}$ | 150 | 400 | 60 |

*All the etching reactions were carried out to room temperature.

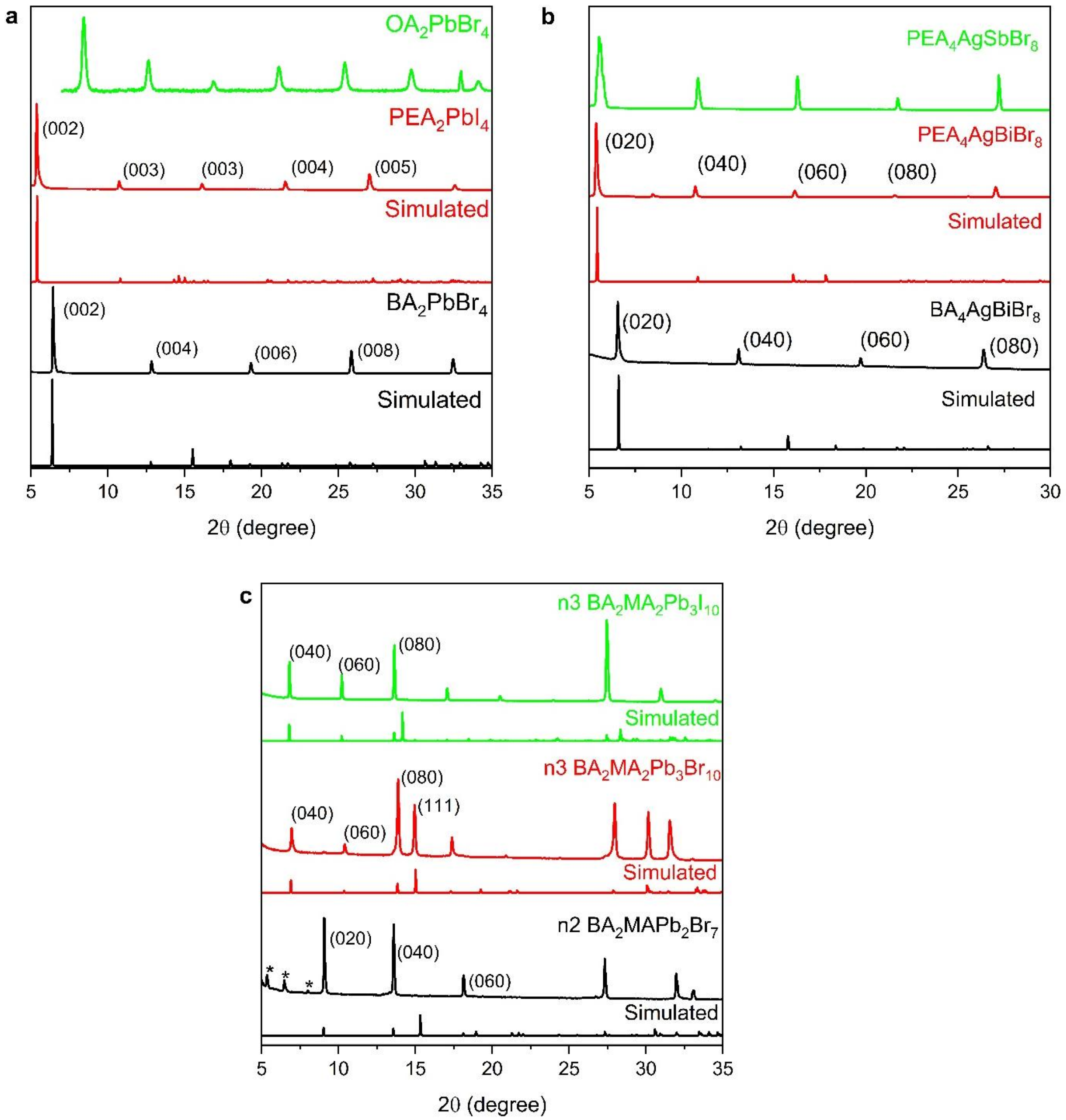

**Figure S15 | XRD patterns of various etched perovskite microcrystals along with the standard simulated patterns.** **a** $BA_2PbBr_4$[7], $PEA_2PbI_4$[8], and $OA_2PbBr_4$. **b** $BA_4AgBiBr_8$[9], $PEA_4AgBiBr_8$[10], and $PEA_4AgSbBr_8$. **c** n2 $BA_2MAPb_2Br_7$[11], n3 $BA_2MA_2Pb_3Br_{10}$[12], and n3 $BA_2MA_2Pb_3I_{10}$[13].

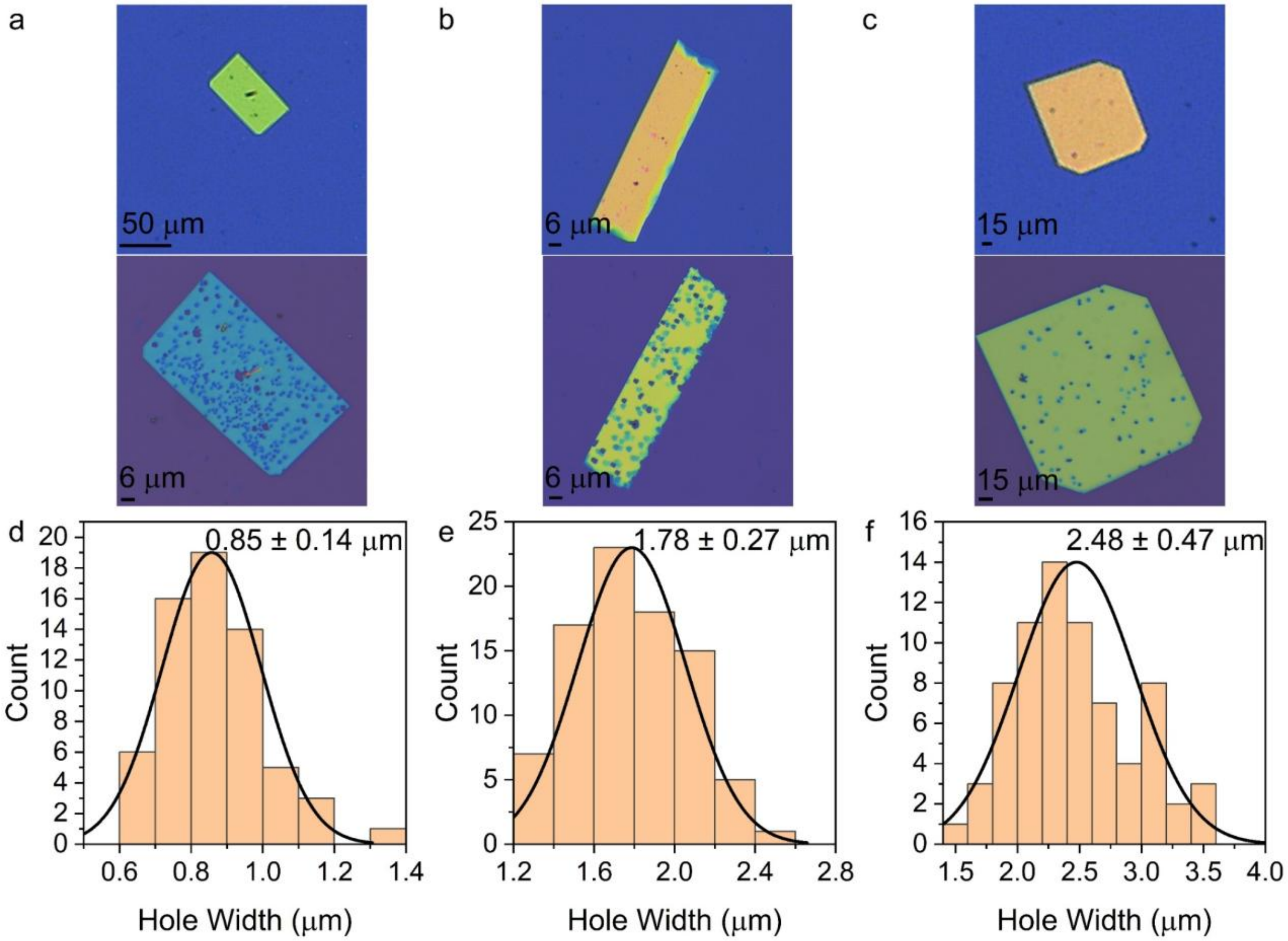

**Figure S16 | Etching of $PEA_2PbBr_4$ microcrystals using alternative antisolvents. a-c** Optical BF images of pre-etched and after etching $PEA_2PbBr_4$ microcrystals at room temperature with etching solution containing acetonitrile as solvent and ethyl acetate, diethyl ether, and n-butanol as anti-solvent, respectively. **d-f** Hole size distribution plots corresponding to etched microcrystals in **a**, **b**, and **c**, respectively.

**Table S3:** Perovskites and the etching solution and etching reaction duration corresponding to the optical images in **Figure S16**.

| Sample (anti-solvent used) | Volume of Acetonitrile (µL) | Volume of anti-solvent (µL) | Etching Reaction Duration (seconds) |
|---|---|---|---|
| **a** (ethyl acetate) | 50 | 450 | 60 |
| **b** (diethyl ether) | 50 | 450 | 60 |
| **c** (n-butanol) | 30 | 470 | 60 |

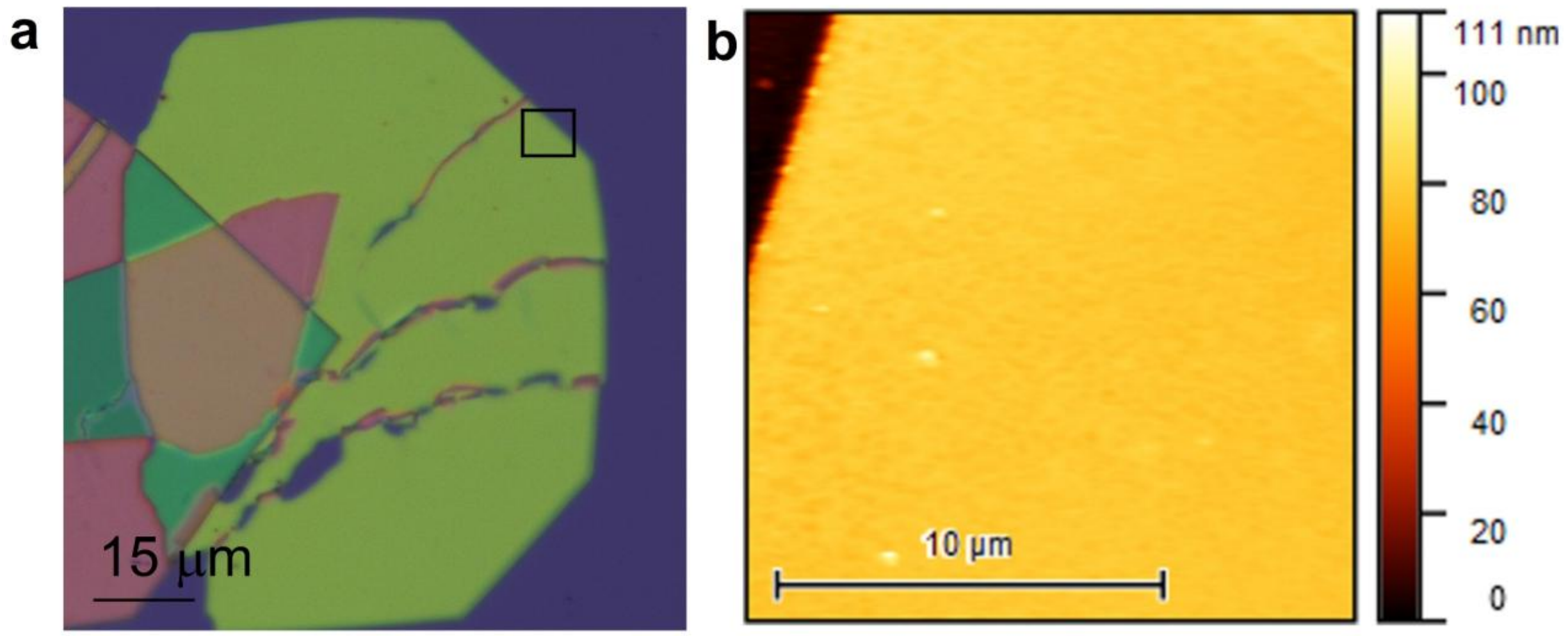

**Figure S17 | Optical and AFM characterization of pristine $PEA_2PbBr_4$ microcrystals. a** Optical BF image of pristine $PEA_2PbBr_4$ microcrystal. **b** AFM image for the marked region in **a**.

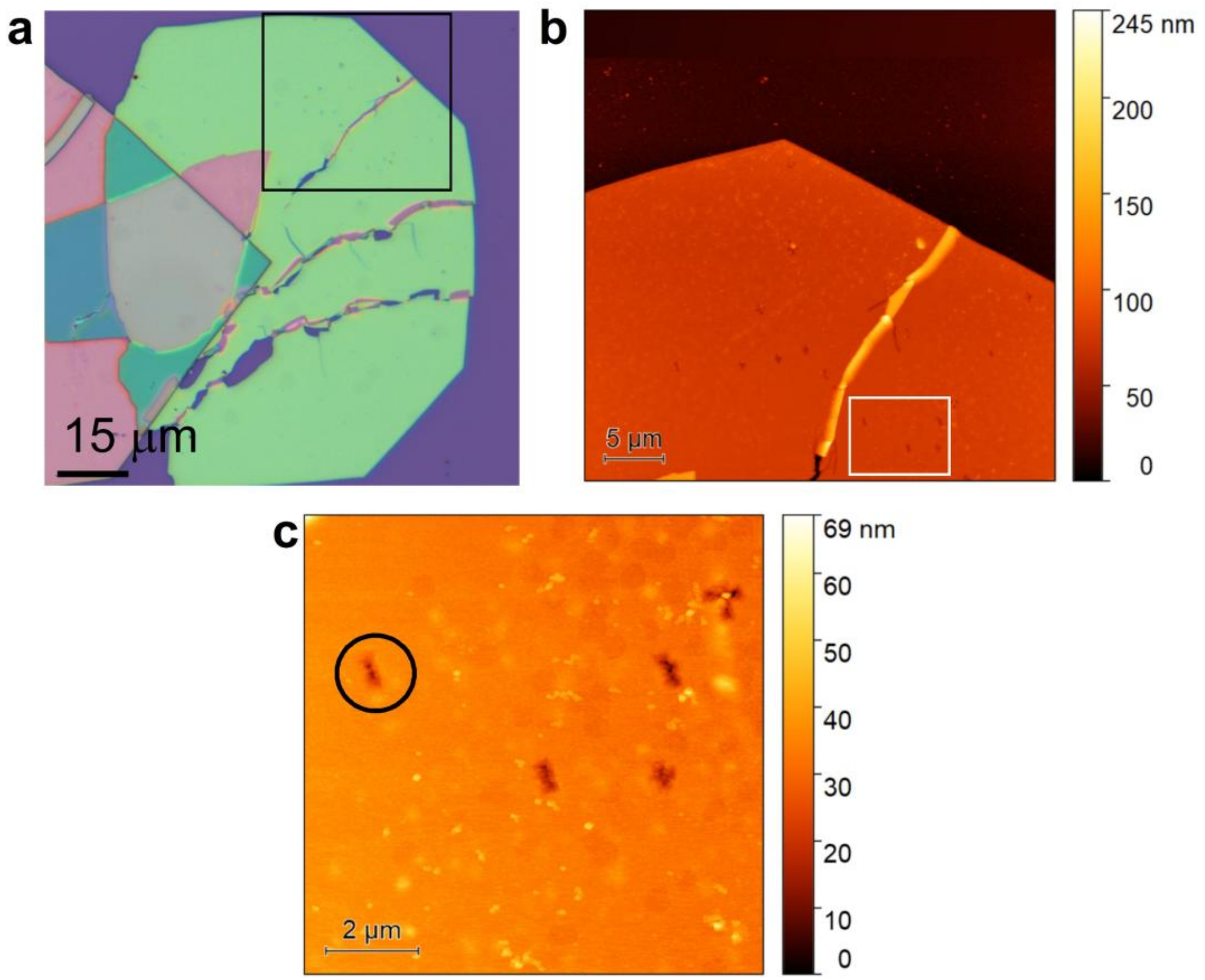

**Figure S18 | Optical and AFM characterization of $PEA_2PbBr_4$ cavities after 2 min of etching.** **a** Optical BF image of $PEA_2PbBr_4$ microcrystal after etching for 2 min at room temperature. **b** AFM image for the marked region in **a**. **c** High-resolution AFM image of region marked in **b**. The selected cavity monitored for cavity evolution is marked in circle in **c**.

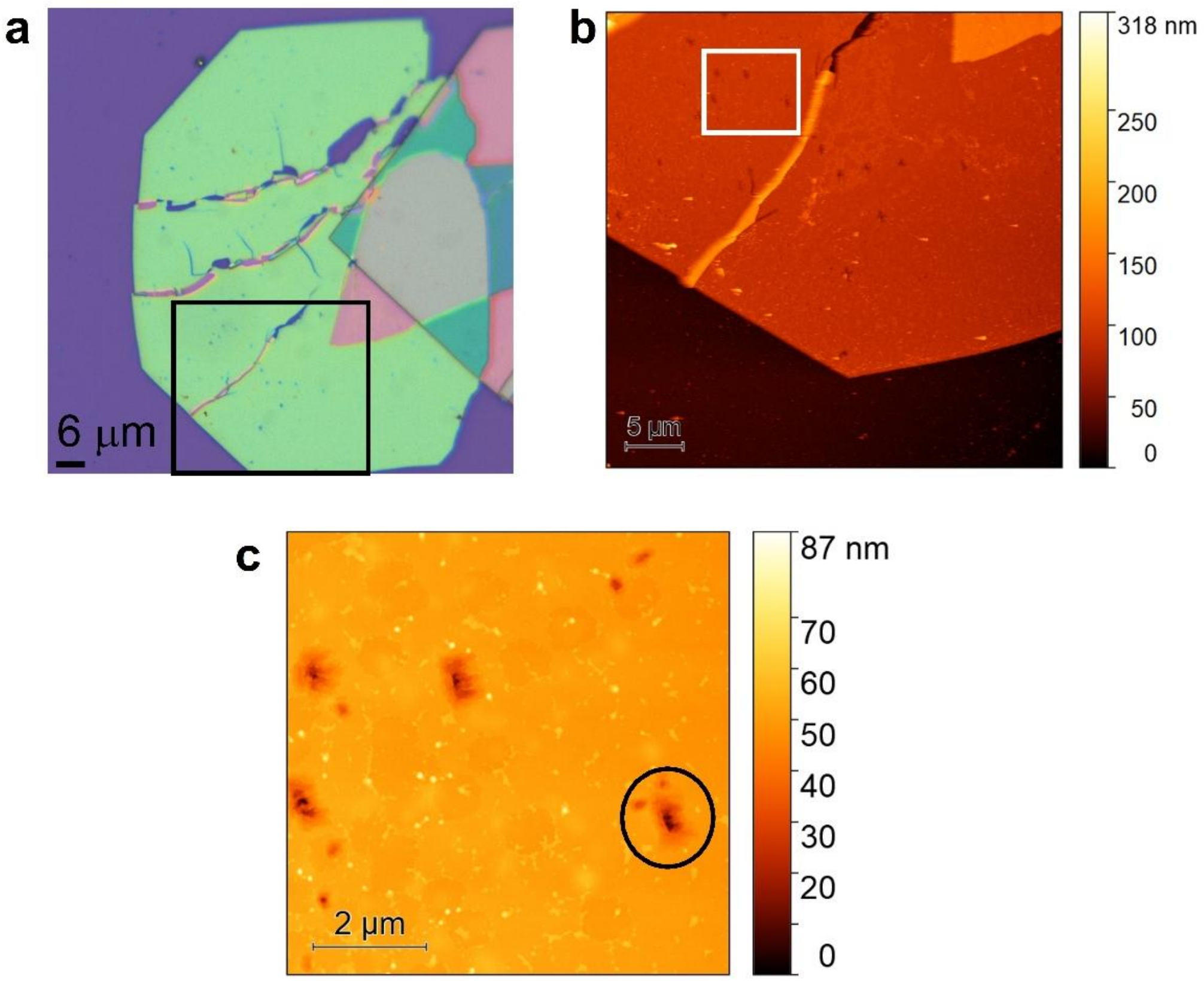

**Figure S19 | Optical and AFM characterization of $PEA_2PbBr_4$ cavities after 4 min of etching. a** Optical BF image of $PEA_2PbBr_4$ microcrystal after etching for 4 min at room temperature. **b** AFM image for the marked region in **a**. **c** High-resolution AFM image of region marked in **b**. The selected cavity monitored for cavity evolution is marked in circle in **c**.

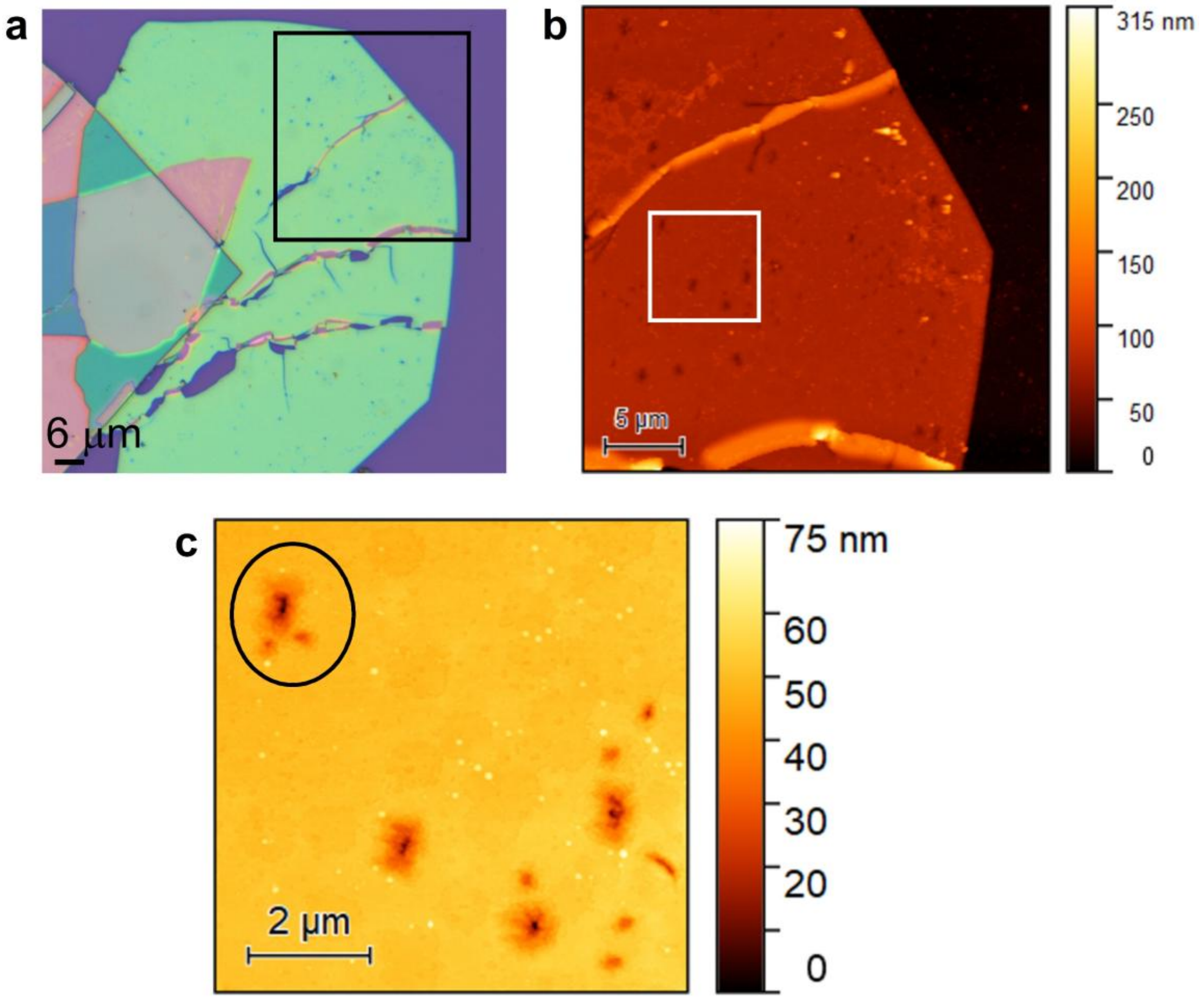

**Figure S20 | Optical and AFM characterization of $PEA_2PbBr_4$ cavities after 6 min of etching.** **a** Optical BF image of $PEA_2PbBr_4$ microcrystal after etching for 6 min at room temperature. **b** AFM image for the marked region in **a**. **c** High-resolution AFM image of region marked in **b**. The selected cavity monitored for cavity evolution is marked in circle in **c**.

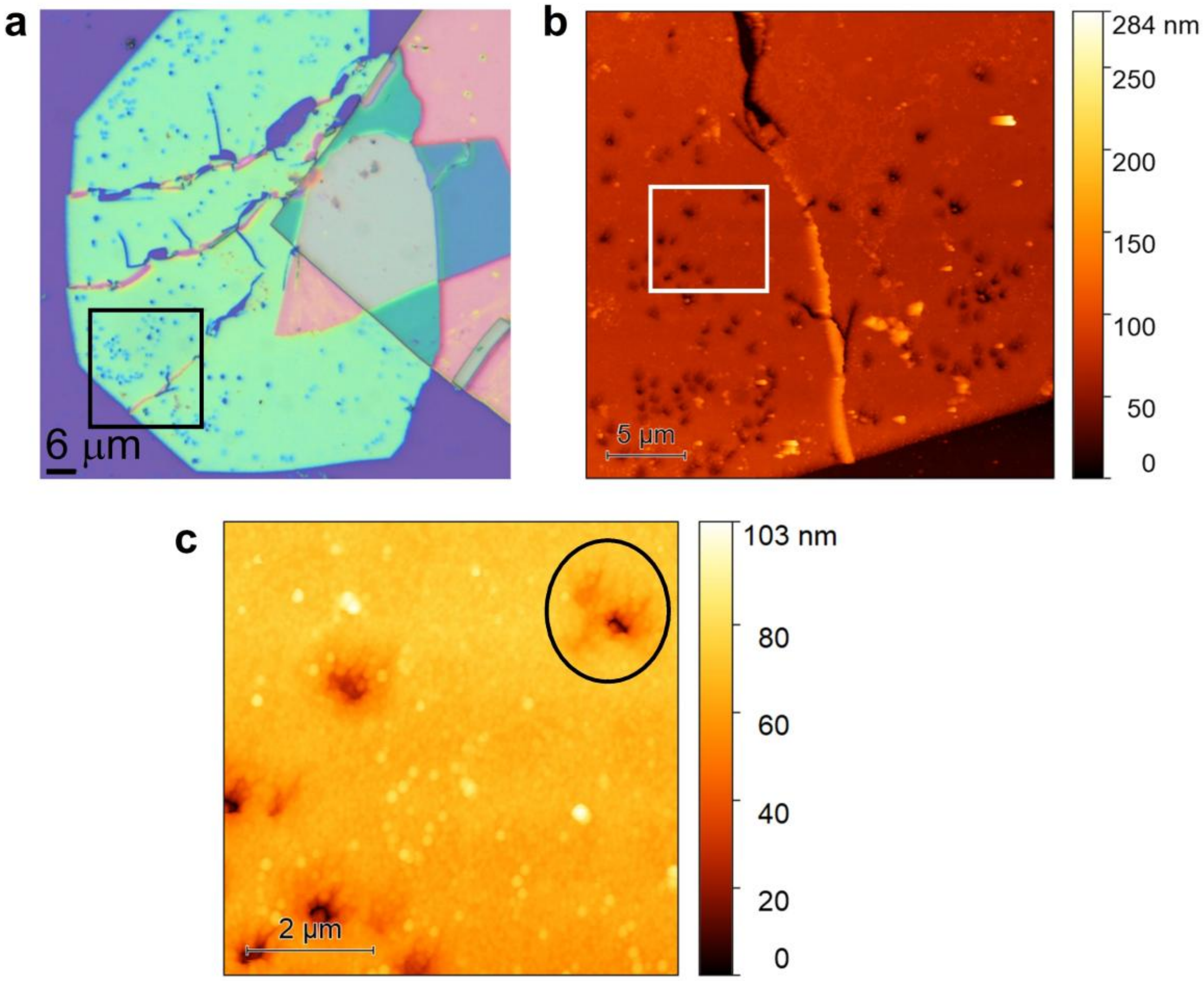

**Figure S21 | Optical and AFM characterization of $PEA_2PbBr_4$ cavities after 11 min of etching. a** Optical BF image of $PEA_2PbBr_4$ microcrystal after etching for 11 min at room temperature. **b** AFM image for the marked region in **a**. **c** High-resolution AFM image of region marked in **b**. The selected cavity monitored for cavity evolution is marked in circle in **c**.

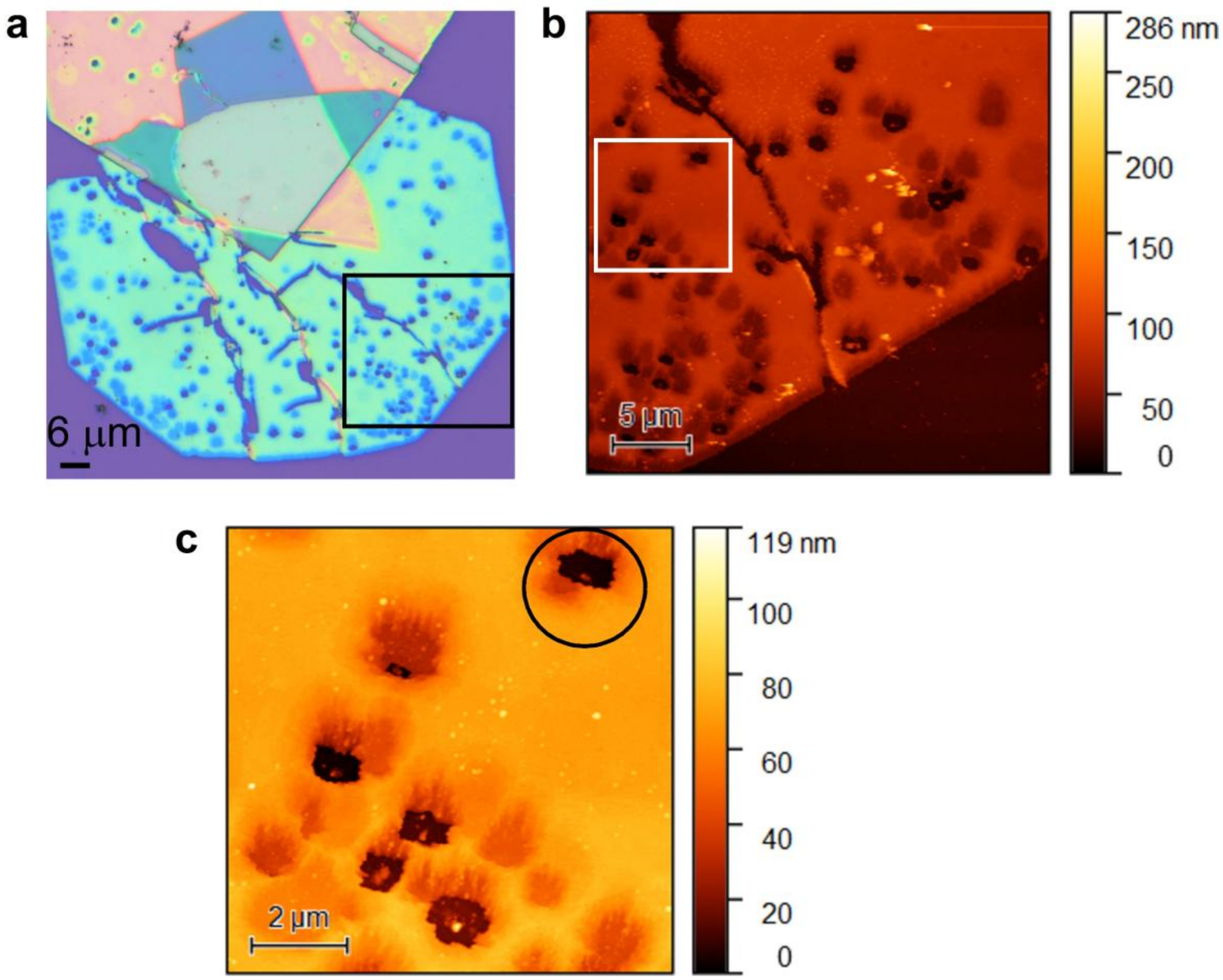

**Figure S22 | Optical and AFM characterization of $PEA_2PbBr_4$ cavities after 30 min of etching.** **a** Optical BF image of $PEA_2PbBr_4$ microcrystal after etching for 30 min at room temperature. **b** AFM image for the marked region in **a**. **c** High-resolution AFM image of region marked in **b**. The selected cavity monitored for cavity evolution is marked in circle in **c**.

**Table S4:** Simulated and experimental cavity dimensions after etching for different times.

| Etching Time (min) | Experimental | | | Simulated | | |
|---|---|---|---|---|---|---|
| | Depth (nm) | Width (µm) | Length (µm) | Depth (nm) | Width (µm) | Length (µm) |
| 2 | 23.4 | 0.50 | 0.91 | 22.3 ± 1.0 | 0.48 ± 0.03 | 0.67 ± 0.04 |
| 4 | 37.1 | 0.64 | 1.07 | 34.2 ± 1.2 | 0.60 ± 0.02 | 0.94 ± 0.05 |
| 6 | 52.0 | 0.79 | 1.15 | 45.2 ± 1.4 | 0.68 ± 0.02 | 1.06 ± 0.05 |
| 11 | 67.4 | 0.95 | 1.41 | 67.6 ± 2.8 | 0.85 ± 0.03 | 1.25 ± 0.06 |
| 30 | 75.0 | 1.36 | 1.91 | 75.0 | 1.21 ± 0.03 | 1.67 ± 0.07 |

**Table S4** compares the simulated depth, width and length with the corresponding AFM-derived measurements. The simulated depth increases from 22 nm after 2 min etching to 67 nm after 11 min etching, eventually reaching the 75 nm substrate boundary at 30 min. These simulated values closely reproduce the experimentally measured depths of 23 nm after 2 min etching, 67 after 11 min and 75 nm at substrate after 30 min. The simulated lateral width along [010] also agrees closely with the experimental measurements, with deviations not exceeding 0.1 µm across the investigated etching times. The simulated cavity length along [100] follows the experimentally observed overall increase, reaching 1.67 µm at 30 min, consistent with the measured merged-cavity length, with deviations less than 0.3 μm at investigated etching times.

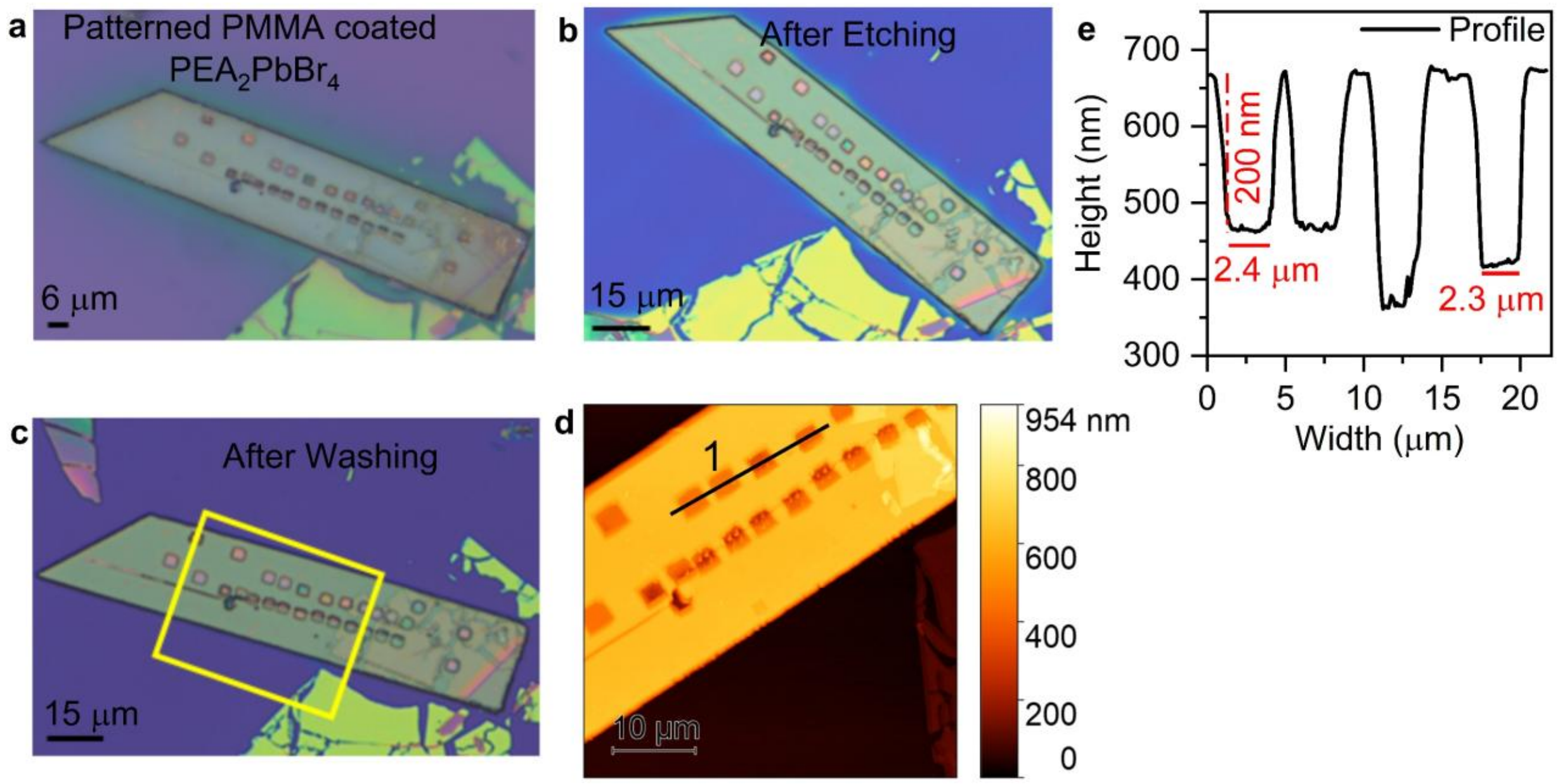

**Figure S23 | Selective etching of $PEA_2PbBr_4$ microcrystals through patterned PMMA masks.** **a** Optical BF image of a $PEA_2PbBr_4$ microcrystal coated with square-hole-patterned PMMA. **b** Optical BF image after etching for 30 s. **c** Optical BF image after removal of the PMMA coating by washing with chlorobenzene. **d** AFM image of the selected region in **c**. **e** Height profile extracted along line 1 in **d**.

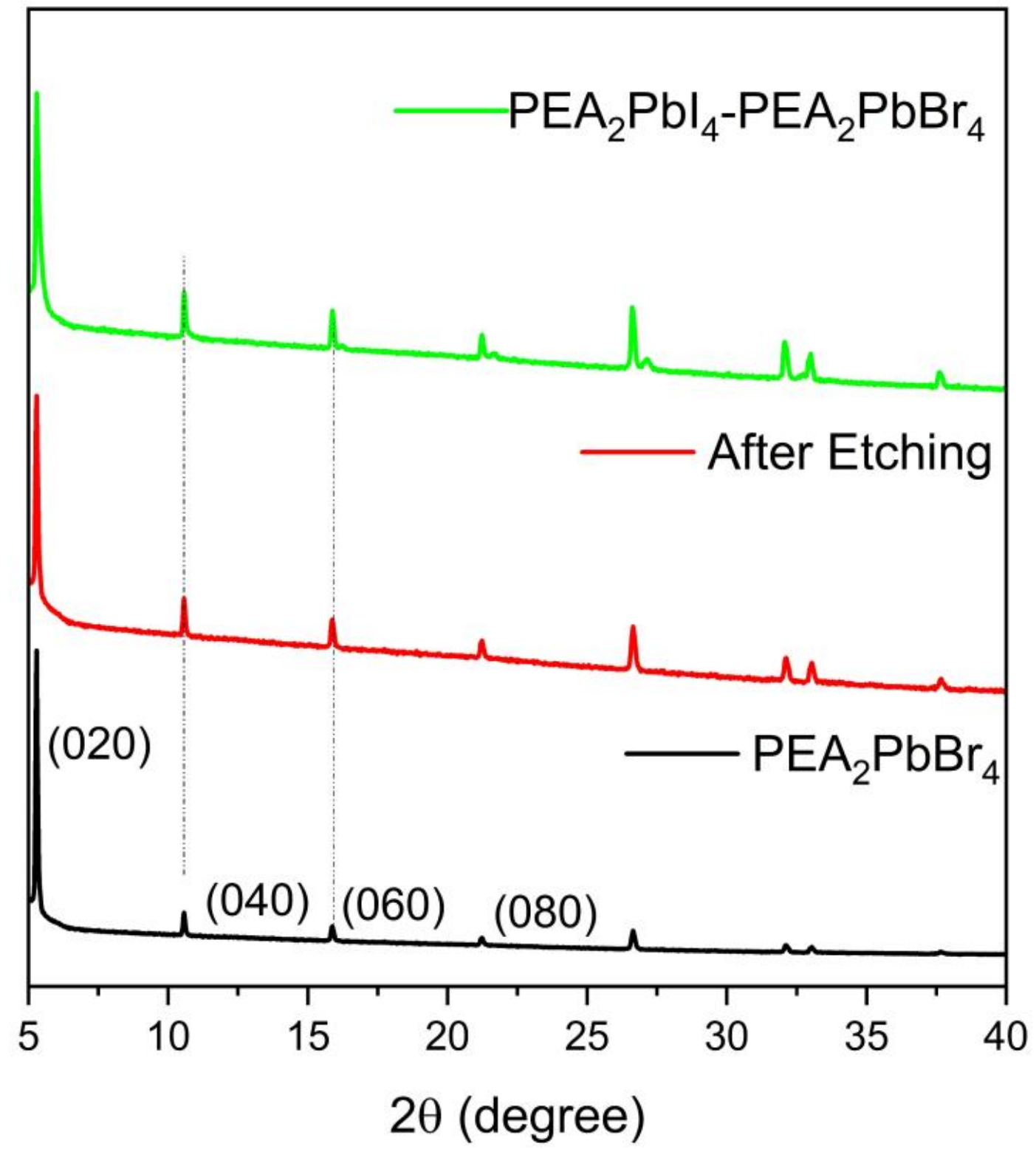

**Figure S24 | XRD characterization of $PEA_2PbI_4$-$PEA_2PbBr_4$ heterostructures.** XRD pattern of $PEA_2PbI_4$-$PEA_2PbBr_4$ heterostructure compared with pure pre-etched and after etching $PEA_2PbBr_4$.

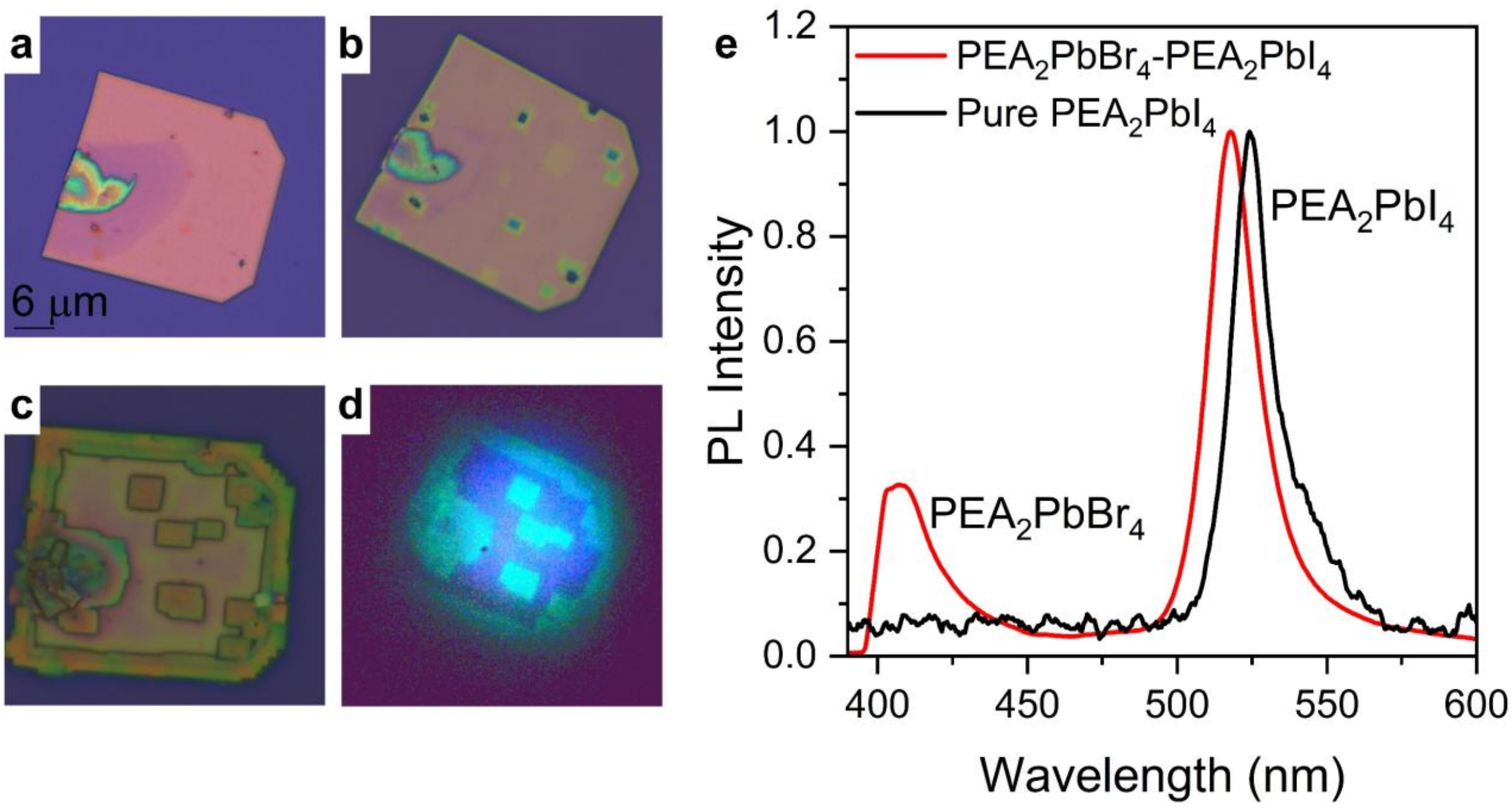

**Figure S25 | Optical and PL characterization of $PEA_2PbI_4$–$PEA_2PbBr_4$ heterostructures.** Optical BF images of $PEA_2PbBr_4$ microcrystal: **a** pre-etched, **b** after etching, **c** $PEA_2PbI_4$-$PEA_2PbBr_4$ heterostructure, and **d** corresponding PL image of heterostructure. **e** PL spectra of $PEA_2PbI_4$-$PEA_2PbBr_4$ heterostructure compared with pure $PEA_2PbI_4$.

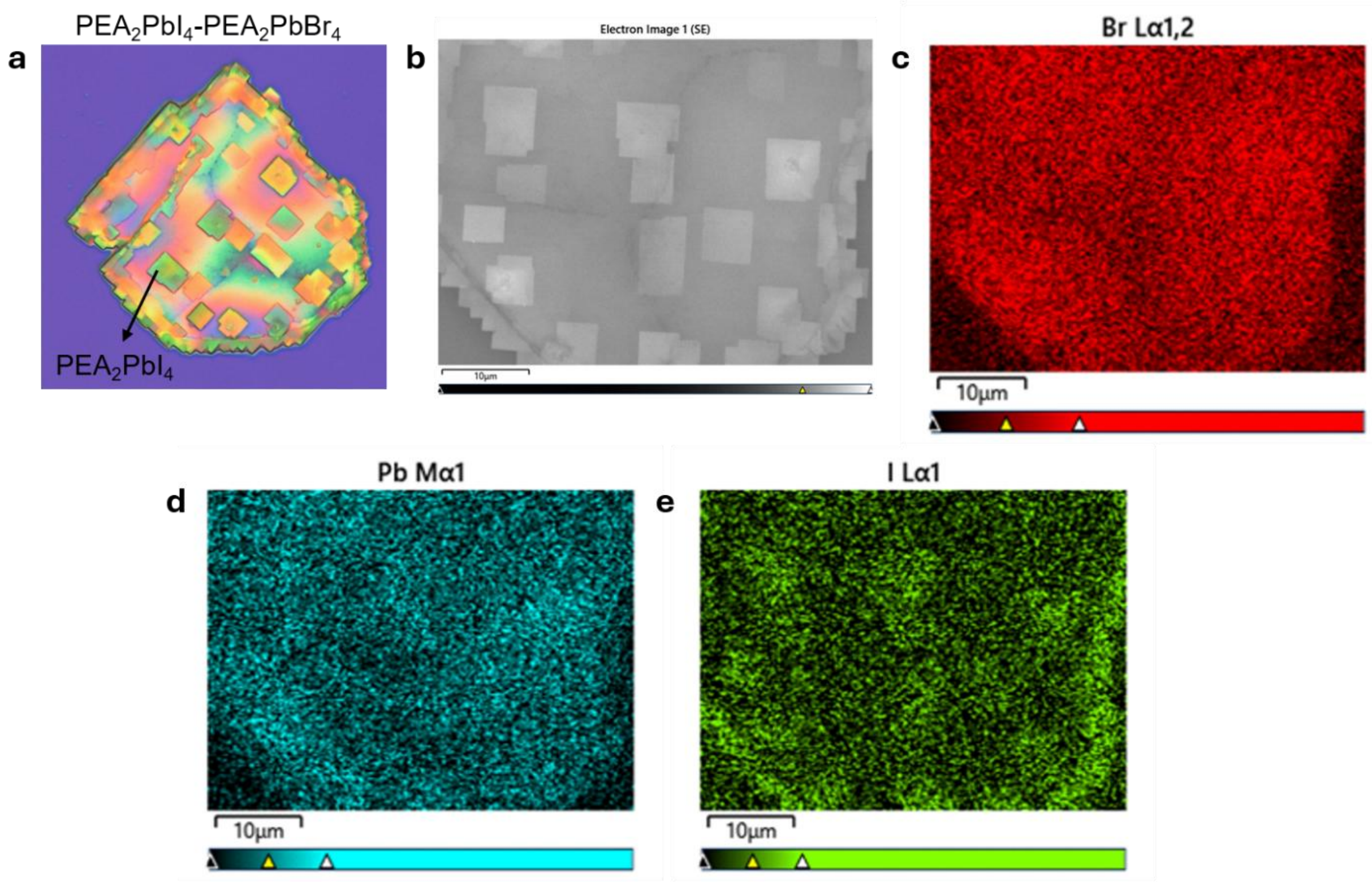

**Figure S26 | Elemental mapping of $PEA_2PbI_4$–$PEA_2PbBr_4$ heterostructures. a** Optical BF image of $PEA_2PbI_4$-$PEA_2PbBr_4$ heterostructure microcrystal. **b** The electronic image of the selected region. **c**, **d**, and **e** the elemental distribution maps for Br, Pb and I, respectively.

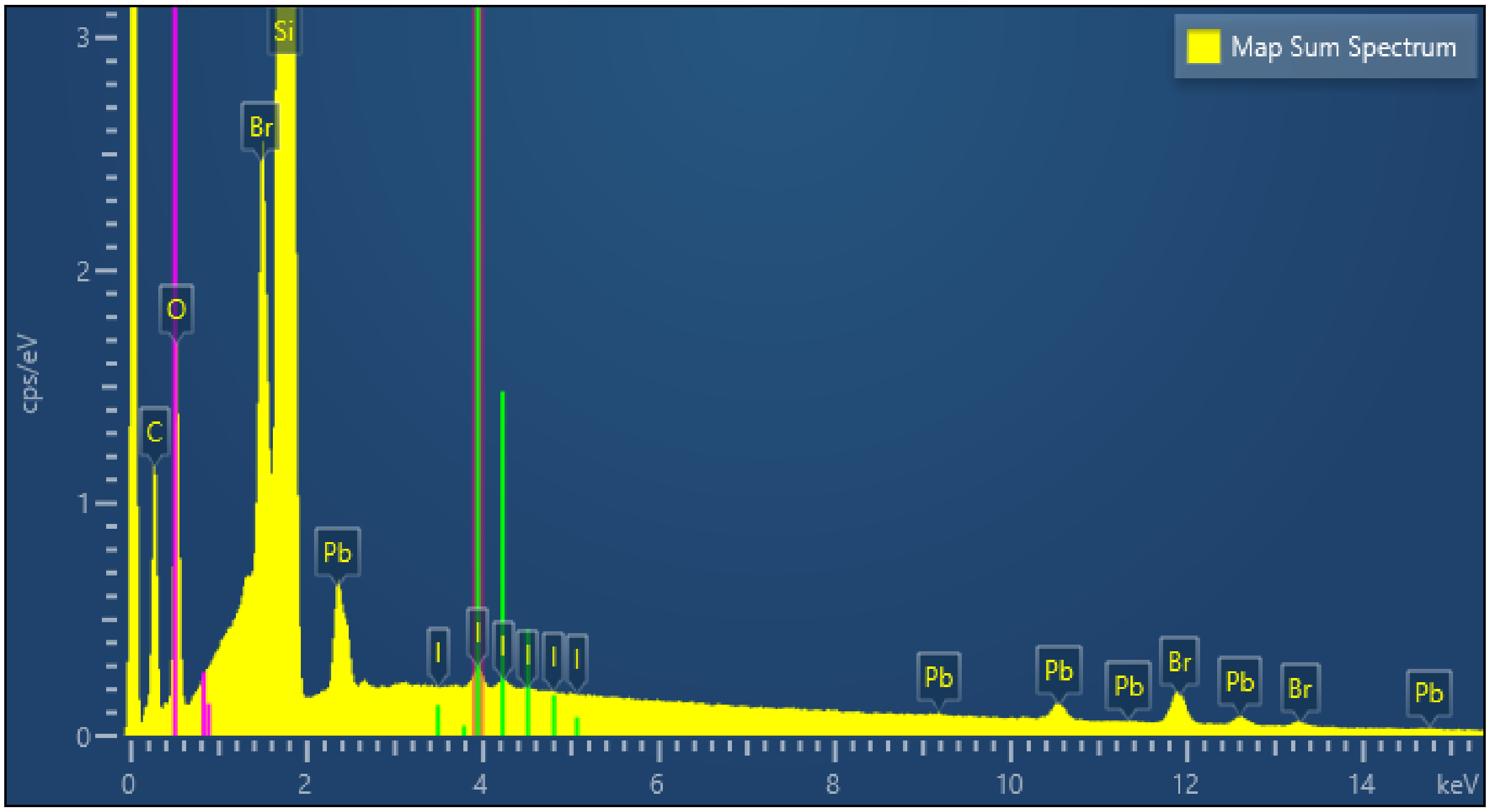

**Figure S27 | Summed EDS spectrum.** The map sum spectrum corresponding to the SEM-EDS mapping in **Figure S26**.

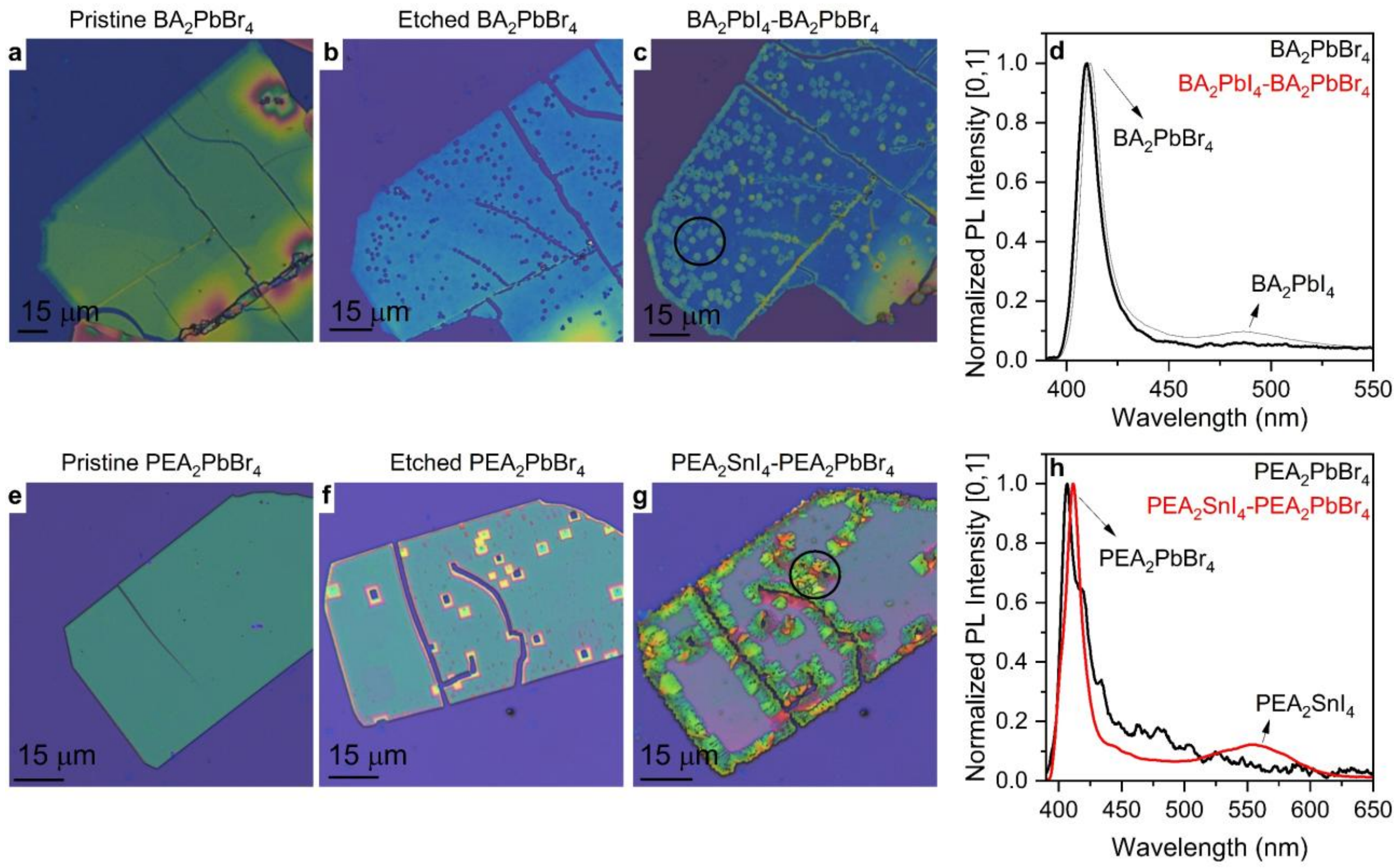

**Figure S28 | Optical and PL characterization of additional perovskite heterostructures. a**, **b**, and **c** Optical BF images of pristine $BA_2PbBr_4$, etched $BA_2PbBr_4$, and $BA_2PbI_4$-$BA_2PbBr_4$ heterostructures. **d** The PL spectra pristine $BA_2PbBr_4$ and $BA_2PbI_4$-$BA_2PbBr_4$ heterostructures. **e**, **f**, and **g** Optical BF images of pristine $PEA_2PbBr_4$, etched $PEA_2PbBr_4$, and $PEA_2SnI_4$-$PEA_2PbBr_4$ heterostructures. **h** The PL spectra pristine $PEA_2PbBr_4$ and $PEA_2SnI_4$-$PEA_2PbBr_4$ heterostructures. Circled spots were used for PL spectra collection.

## References

1. Pan, D. *et al.* Deterministic fabrication of arbitrary vertical heterostructures of two-dimensional Ruddlesden–Popper halide perovskites. *Nat. Nanotechnol.* **16**, 159–165 (2021).
2. Zhang, S. *et al.* Mosaic lateral heterostructures in two-dimensional perovskite. *Nature* **649**, 612–620 (2026).
3. Borreani, M., Saini, S. K., Schleusener, A. & Krahne, R. Direct Growth of Rectangular 2D Layered Metal-Halide Perovskite Microcrystal Photonic Cavities on Functional Substrates. *Adv. Opt. Mater.* **13**, e01276 (2025).
4. Klinger, M. More features, more tools, more CrysTBox. *J. Appl. Crystallogr.* **50**, 1226–1234 (2017).
5. Flidr, J., Huang, Y.-C., Newton, T. A. & Hines, M. A. Extracting site-specific reaction rates from steady state surface morphologies: Kinetic Monte Carlo simulations of aqueous Si(111) etching. *J. Chem. Phys.* **108**, 5542–5553 (1998).
6. Meakin, P. & Rosso, K. M. Simple kinetic Monte Carlo models for dissolution pitting induced by crystal defects. *J. Chem. Phys.* **129**, 204106 (2008).
7. Mitzi, D. B. Synthesis, Crystal Structure, and Optical and Thermal Properties of $(C_4H_9NH_3)_2MI_4$ (M = Ge, Sn, Pb). *Chem. Mater.* **8**, 791–800 (1996).
8. Du, K. *et al.* Two-Dimensional Lead(II) Halide-Based Hybrid Perovskites Templated by Acene Alkylamines: Crystal Structures, Optical Properties, and Piezoelectricity. *Inorg. Chem.* **56**, 9291–9302 (2017).
9. Connor, B. A., Leppert, L., Smith, M. D., Neaton, J. B. & Karunadasa, H. I. Layered Halide Double Perovskites: Dimensional Reduction of $Cs_2AgBiBr_6$. *J. Am. Chem. Soc.* **140**, 5235–5240 (2018).

10. Schmitz, F. *et al.* Large Cation Engineering in Two-Dimensional Silver–Bismuth Bromide Double Perovskites. *Chem. Mater.* **33**, 4688–4700 (2021).

11. Li, L. *et al.* Two-Dimensional Hybrid Perovskite-Type Ferroelectric for Highly Polarization-Sensitive Shortwave Photodetection. *J. Am. Chem. Soc.* **141**, 2623–2629 (2019).

12. Li, L. *et al.* Tailored Engineering of an Unusual $(C_4H_9NH_3)_2(CH_3NH_3)_2Pb_3Br_{10}$ Two-Dimensional Multilayered Perovskite Ferroelectric for a High-Performance Photodetector. *Angew. Chem. Int. Ed.* **56**, 12150–12154 (2017).

13. Stoumpos, C. C. *et al.* Ruddlesden–Popper Hybrid Lead Iodide Perovskite 2D Homologous Semiconductors. *Chem. Mater.* **28**, 2852–2867 (2016).